\documentstyle[aaspp4,12pt]{article}
\begin{document}

\title{Star Formation Histories from HST Color-Magnitude Diagrams of Six Fields of the LMC\footnote{Based on observations made with the NASA/ESA {\it Hubble Space Telescope}, obtained at the Space Telescope Science Institute, which is operated by the Association of Universities for Research in Astronomy, Inc. under NASA contract NAS5-26555.}}
\author{Knut A.G. Olsen}
\affil{Cerro Tololo Inter-American Observatory, Casilla 603, La Serena, Chile}

\begin{abstract}
We present results on the analysis of background field stars found in {\it HST} WFPC2 observations of six of the old globular clusters of the Large Magellanic Cloud.  Treated as contaminants by the globular cluster analysts, we produce $V-I,V$ color-magnitude diagrams of the field stars and use them to explore the LMC's star formation history.  The photometry approaches $V\sim26$, well below the turnoff of an ancient ($\sim$14 Gyr) LMC population of stars.  The field star CMDs are generally characterized by an upper main sequence broadened by stellar evolution, an old red giant branch, a prominent red clump, and an unevolved lower main sequence.  The CMDs also contain a few visual differences, the most obvious of which is the smeared appearance of the NGC 1916 field caused by heavy differential reddening.  More subtly, the base of the subgiant branch near the old turnoff appears extended in $V$ and the red giant branch appears broad in $V-I$ in four of the fields, but not in the NGC 1754 field.
We use a maximum-likelihood technique to fit model CMDs drawn from Bertelli et al.\ (1994) isochrones to the observed CMDs.  We constrain the models by the age-metallicity relation derived from LMC clusters, test four IMF slopes, and fit for the reddening, distance modulus, and star formation rate.
We find that we can just resolve structure in SFR($t$) with time steps of $\sim$0.15 in log age, implying a resolution of $\sim4\times10^8$ years at an age of 1 Gyr.  For a Salpeter IMF, our derived star formation history for the NGC 1754 field is characterized by an enhanced star formation rate over the past 4 Gyr, qualitatively resembling that derived by others for a variety of LMC fields.  The remaining four fields, however, appear to have had high levels of star formation activity as long as 5$-$8 Gyr ago; these fields lie in the LMC Bar while the NGC 1754 field lies in the disk, suggesting that the inner regions of the LMC contain significantly more older stars than the outer regions.  Examining the residuals of the models and observations, we find that the old red giant branches of the models provide a poor fit to the observations, which suggests an error in the model isochrones.  The effect of the disagreement appears to be to underestimate the contribution of the old population. 

\end{abstract}

\section{Introduction}

Both because of its proximity and its unique nature, the Large Magellanic Cloud is an excellent laboratory for studying the effects of a variety of processes thought to affect the formation of stars, e.g. interactions with neighboring galaxies, the influence of bar dynamics, stellar feedback, and metallicity of the environment.  Thus, knowledge of the star formation history of the LMC is a step towards understanding the physical origin of stellar populations.  
This star formation history is recorded in the LMC's numerous clusters and field stars.  Because clusters are relatively easy to study, measuring their age and metallicity distribution is a natural way to study the star formation history of the LMC.  Researchers have shown that the LMC's clusters are predominantly younger than 4 Gyr in age, with only a small number of old ($\sim$14 Gyr) globular clusters and one cluster, ESO 121+SC03, with an age of $\sim$8 Gyr (Mateo, Hodge, \& Schommer 1986).  

While the cluster results suggest that the LMC has formed the bulk of its stars over the past 4 Gyr, recent observations of LMC field stars taken with the {\it Hubble Space Telescope} imply that stars formed even during the lull in cluster formation, assuming a standard Salpeter IMF (Holtzman et al.\ 1997, hereafter H97, Geha et al.\ 1998).  These provocative results open a number of questions, such as what might cause a switch between cluster and field star formation modes, and indicate that field stars need to be accounted for in a complete description of the LMC's star formation history.  Moreover, as a complete record of star formation is available in a single pointing and field stars have an almost continuous age distribution, field observations appear ideally suited for identifying major star formation events and for mapping spatial variations in the star formation history, both keys for understanding the physical origin of stellar populations.

The clear disadvantage of using field stars rather than clusters to study the star formation history, however, is that the information they hold is difficult to extract.  This difficulty was recognized by Schwarzschild (1958), who wrote: ``The random sample represented by the nearby stars certainly includes a large range of masses, possibly some variety in initial composition, and most likely a great mixture of stellar ages.  A great variety of evolutionary tracks therefore is involved ... and their disentanglement seems somewhat hopeless.''  
Luckily, modern tools now allow astronomers to attack the problem with brute force on a scale that Schwarzschild could not have done.  Observational facilities such as {\it HST} and a variety of excellent ground-based telescopes gather data efficiently and accurately, while the stellar evolutionary models used in comparison with the color-magnitude diagrams (CMDs) have substantially improved (e.g. Schaller et al.\ 1992, Bertelli et al.\ 1994, hereafter B94, Chaboyer \& Kim 1995, VandenBerg 1998).  Aided by a number of techniques that take advantage of easily available fast computing facilities, work on the star formation histories of the Magellanic Clouds and of other Local Group galaxies is proceeding at a rapid pace (e.g. Gallart et al.\ 1996, Aparicio, Gallart, \& Bertelli 1997,  Dohm-Palmer et al.\ 1998, Tolstoy et al.\ 1998), often with dramatic results (e.g. Smecker-Hane 1997).  

\begin{figure}
\plotone{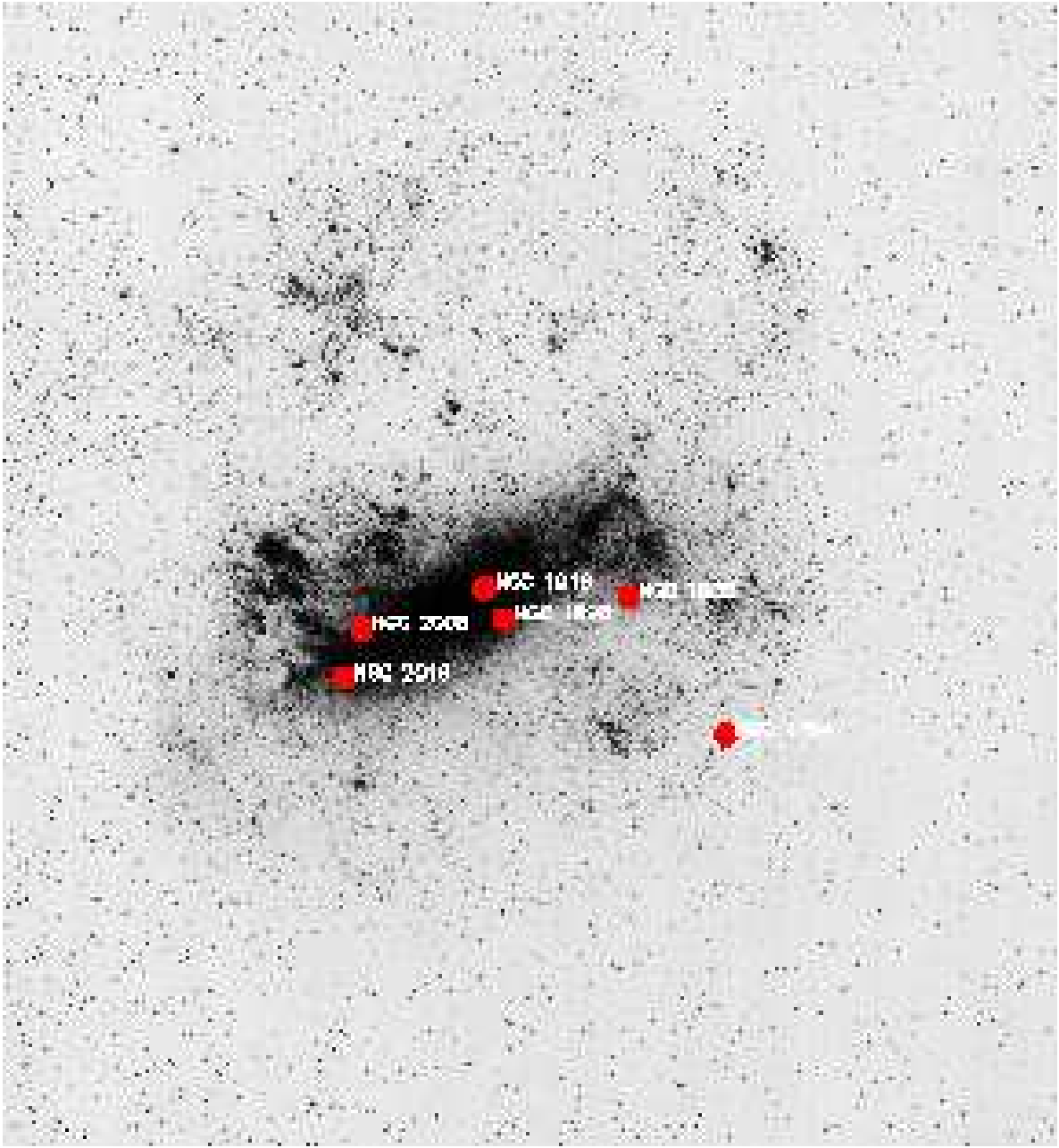}
\caption[Image of the LMC with Inner Globular Clusters]{This image shows the positions of the target LMC fields with respect to the parent galaxy.  Plate courtesy of Harlow Shapley, Harvard College Observatory, Boyden Observatory photograph.}
\label{targets}
\end{figure}

Our present work seeks to add to the growing body of knowledge of the star formation history of the LMC through observations of six fields taken with {\it HST}.  These observations were taken with the primary purpose of producing CMDs of the LMC's inner old globular clusters (Olsen et al.\ 1998, hereafter Paper 1).  However, the images also contain significant numbers of field stars; viewed as contaminants in Paper 1, they are the focus of this paper.  The locations of the fields are shown in Fig.\ \ref{targets}; five of them lie in the Bar while the other lies in the disk.  As the fields are well-separated, the observations are nicely situated for exploring spatial variations in the star formation history.  In particular, we will address the question of whether the Bar and disk have had different star formation histories, as would be expected if the Bar induces star formation or if the event that formed the Bar also triggered a significant star forming episode.
While the majority of published LMC field star studies have avoided the Bar because of the severe effects of photometric crowding, the resolution afforded by {\it HST} enables us to produce 
CMDs which include stars well below the oldest main sequence turn-off even in the Bar.

In the following we briefly describe the observations and reductions in Section 2, referring the bulk of this discussion to Paper 1.  Section 3 describes the method used to extract the star formation histories.  The method, which is based on the work of Dolphin (1997), determines the star formation history by fitting a linear combination of model isochrones to the histogram of points in the CMD (the ``Hess diagram"), accounting for observational errors.  Tests of the method using simulated CMDs are described in Section 4.  Section 5 discusses the application of the method to the observed CMDs and evaluates its success.  We present our conclusions in Section 6. 

\section{Observations and Reductions}
\begin{figure}
\plotone{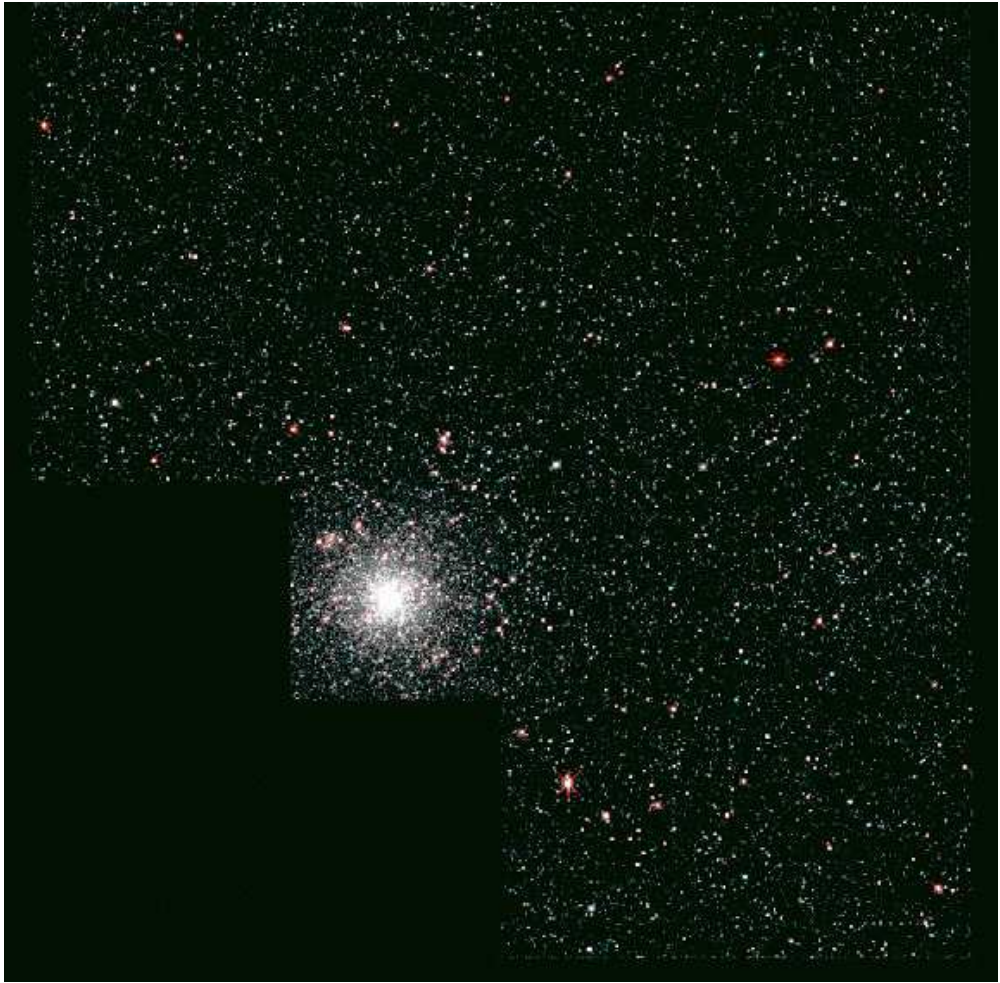}
\caption{``True-color'' WFPC2 mosaic formed from the F555W and F814W frames of NGC 2019.  The WF chips contain mostly field stars and the PC mostly cluster stars, although there is significant overlap.}
\label{image}
\end{figure}

As described in Paper 1, observations of the six LMC globular clusters NGC 1754, NGC 1835, NGC 1898, NGC 1916, NGC 2005, and NGC 2019 were taken during Cycle 5 of {\it HST} using WFPC2 and the F555W and F814W filters.  By taking both long and short exposures through each filter, we were ensured reliable photometry of the brightest stars in the fields while still reaching $V\gtrsim25$.  We centered the clusters on the Planetary Camera, as shown in the sample image (Fig.\ \ref{image}), leaving the Wide Field (WF) chips containing primarily field stars.

Paper 1 fully describes the image reductions, photometry, and production of CMDs.  In summary, cosmic ray rejection, correction for the $y$-dependent CTE effect (Holtzman et al.\ 1995), geometric distortion correction, and bad pixel flagging were applied to the images before performing photometry.  For the photometry, we used version 2.5 of the profile-fitting photometric package DoPHOT (Schechter, Mateo, \& Saha 1993), modified by Eric Deutsch to handle floating-point images.  For the WF chips, aperture corrections were applied as a function of position in the frame using measurements of isolated bright, unsaturated stars taken in the profile-subtracted images.  
In agreement with the report of Whitmore \& Heyer (1997), we found an offset in mean magnitudes between the short- and long-exposure photometry.  We corrected for this effect by adjusting the short-exposure magnitudes to match, on average, those of the long exposures.  Finally, we merged the short- and long-exposure lists of photometry, transformed the magnitudes from the WFPC2 system to Johnson $V$/Kron-Cousins $I$, and produced CMDs by matching stars with measurements in both $V$ and $I$ by position.  Table \ref{wfphot}a contains a sample portion of the photometry, the full version of which is published in the electronic version of the {\it Astronomical Journal}.

\begin{figure}
\plotone{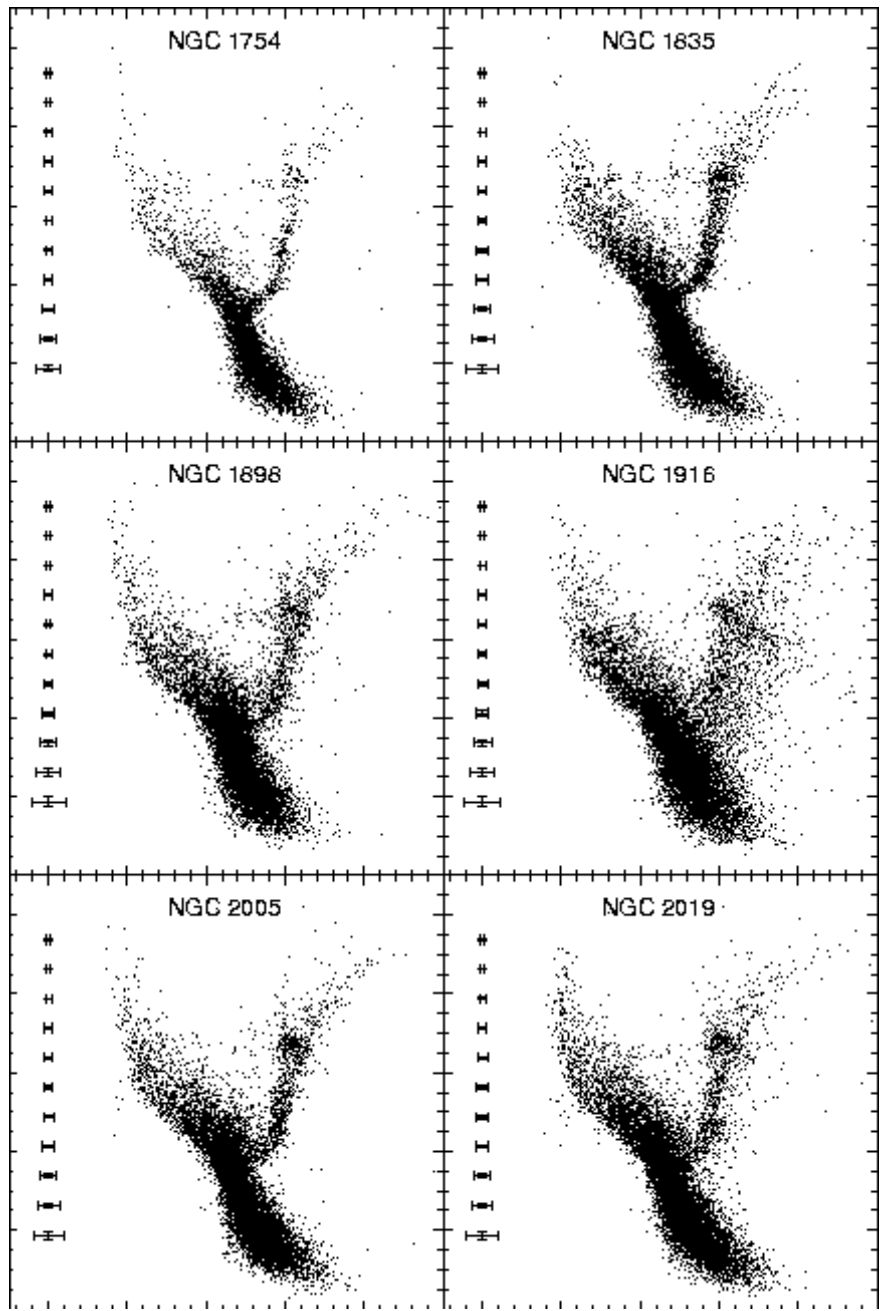}
\caption{Color-magnitude diagrams of the combined WF chips of each field, restricted to those stars with DoPHOT object type 1 (see discussion in Olsen et al.\ 1998).  Each CMD is roughly characterized by a broad upper main sequence, a lower main sequence extending to $V\sim25$, and an old red giant branch.  The smeared appearance of the NGC 1916 field is evidence of strong differential reddening.}
\label{rawcmd}
\end{figure}

Fig.\ \ref{rawcmd} shows the CMDs obtained from the combined WF frames of each field.  The main features of the CMDs are a lower main sequence (MS) seen to $V\sim25.5$, a broad upper MS, a prominent red clump (RC) containing intermediate-age horizontal branch (HB) stars, and an old red giant branch (RGB).  The stars falling to the blue of the main sequence envelope at $V\sim21$ in the NGC 1835, NGC 2005, and NGC 2019 fields could be blue HB stars; however, most old, metal-poor HB stars will be difficult to distinguish from the broad upper main sequence.  While the old red giant branch in the NGC 1754 field forms a tight sequence, in the other fields (excluding NGC 1916) the giant and subgiant branches appear broader, suggesting a range of turnoff ages.  The NGC 1916 field CMD looks remarkably different from the others, having a much broader lower MS and RGB and a stretched-out RC.  As was discussed in Paper 1, this appearance is most likely the result of strong differential reddening, clear evidence for which is seen in ground-based images of the region (A. Walker, personal communication).  While we cannot rule out a contribution from significant depth in the field stars local to NGC 1916, the simplest hypothesis is that differential reddening entirely accounts for the anomalous appearance of the NGC 1916 field CMD.

\begin{figure}
\plotone{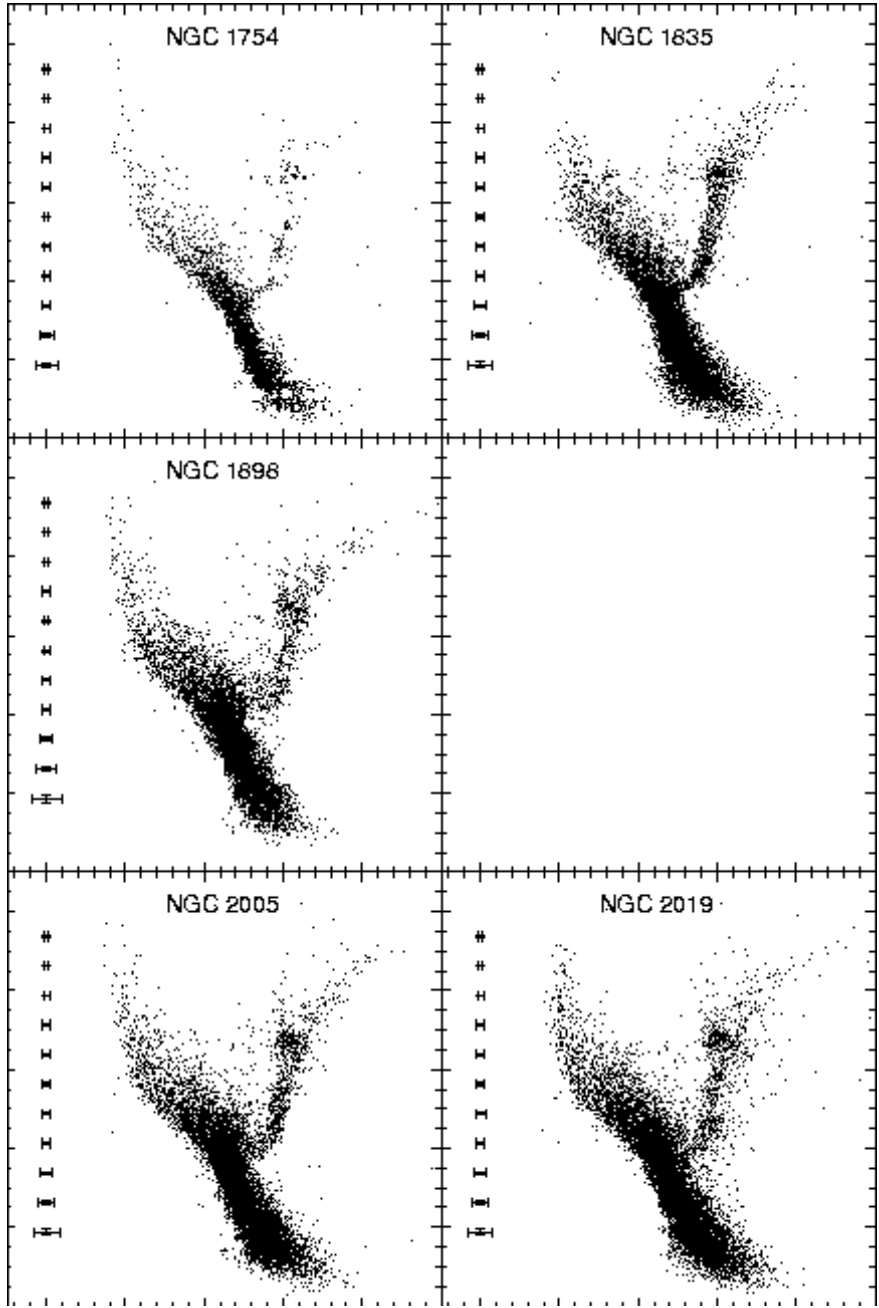}
\caption{Field star CMDs after statistical subtraction of cluster stars.  The subtraction has removed significant portions of the old RGBs, particularly in the NGC 1754 and NGC 1898 fields.  The subtraction is not perfect, in some cases carving out regions of the CMDs.}
\label{subcmd}
\end{figure}

Some of the old RGB and lower MS stars are expected to be globular cluster stars spilling over into the WF chips.  From the estimated core and tidal radii of the clusters (Paper 1), we predict the numbers of contaminants listed in Table \ref{contam}.  In Paper 1, we described using these estimates to iteratively subtract field stars from the cluster CMDs and vice-versa.  Fig.\ \ref{subcmd} shows the field star CMDs after the cluster stars have been removed.  The NGC 1754 and NGC 1898 field CMDs have been altered significantly, with the NGC 1754 field CMD losing most of its RGB, while the others appear relatively unchanged.  The subtraction was not perfect, as some regions of the CMDs have a choppy appearance; it appears that the large number of contaminants present in some of the fields brings out the faults of the subtraction technique.  We did not find this problem when subtracting field stars from the cluster CMDs, as there are fewer contaminating field stars in the PC.  Moreover, the estimated number of contaminants in the fields depends strongly on the somewhat uncertain tidal radii of the King model fits to the cluster profiles.  For these reasons, we decided it was easier to model the field star CMDs without correcting for contamination from the clusters; all of the star formation histories we derive thus include a contribution from the globular clusters.

\section{Method for Determining the Star Formation Histories}
The study of the star formation history of the LMC using field star CMDs has a history dating to more than 20 years ago, the biggest developments of which are summarized by Gallagher et al. (1996).  The earliest analyses were conducted by visual examination of the main features of the CMDs (the upper main sequence, RGB, and asymptotic giant branch) and by comparing the CMDs to fiducial sequences of clusters of known age and metallicity.  Bertelli et al. (1992) developed a more sophisticated technique which involved solving for parameters governing the star formation history of their LMC fields.  The solutions were  made by comparing the average density of stars in broad regions of the observed CMDs with predictions from stellar evolutionary models.  Similar techniques have been developed and refined by Gallart et al. (1996), among others, and used to study the star formation histories of Local Group dwarf galaxies.  Recently, H97 and Geha et al. (1998) used the luminosity functions of three LMC fields to fit several model star formation histories.

A common feature of the methods described in the previous paragraph is that they do not use all of the information available in the field star CMDs; they use either selected portions of the CMDs or luminosity functions to extract the star formation histories.  By disregarding the full distribution of colors and magnitudes, those approaches avoid the problems of modelling the uncertain later phases of stellar evolution, but at a cost in age resolution.  Higher resolution is desirable for addressing questions such as the possible effect of interactions on the LMC star formation history and for exploring differences among CMDs taken in different fields.  The theoretical framework for a more detailed comparison of models and observations has been assembled by a number of researchers (e.g. Tolstoy \& Saha 1996, Dolphin 1997, Ng 1998), but has been mostly untested in full capacity against observations.  As our CMDs are well-populated by stars in most of the phases of evolution seen in the models, and we wish to address the question of whether there are differences among the star formation histories they represent, our CMDs provide good motivation and a favorable environment for exploring the potential of a detailed comparison with the models.

\subsection{Principles of the method}
To first approximation, the field star CMDs are a linear combination of isochrones of several ages and metallicities governed by a particular star formation rate as a function of time (SFR($t$)), a universal initial mass function (IMF), an age-metallicity relation (Z($t$)), a single value of the reddening ($E(B-V)$), and a single distance modulus ($(m-M)_\circ$).  Additional factors that may affect the appearance of the CMDs are the presence of binary stars, the spread in Z at each age $t$, variation of the IMF, and the spread in $(m-M)_\circ$ and $E(B-V)$ within the fields.  For this work we ignore these additional factors; the first three are disregarded for the sake of simplicity.  We consider the assumption of uniform distance to be a safe one, as the LMC is known to be thin compared to its distance (Caldwell \& Coulson 1986).  Also, from the tight appearance of the cluster sequences in 5 of the 6 fields (Paper 1), the amount of differential reddening must be small.  For the NGC 1916 field, however, differential reddening is obviously strong; we thus exclude this field from the current analysis.

Given these assumptions, the problem of determining the star formation history is reduced to constructing a model CMD whose SFR($t$) best reproduces the distribution of stars in the observed CMD for a given Z($t$), IMF, $E(B-V)$, and $(m-M)_\circ$.
  One approach to this problem is to construct model CMDs from a small number of template star formation histories and to compare them with the observed CMDs (e.g. Gallart et al.\ 1996, Aparicio et al.\ 1997).
However, in order to take full advantage of the excellent photometry of LMC stars possible with $HST$,  the number of templates we would need to construct would be exceedingly large.  For example, if we wished to extract the star formation history using 10 age bins spanning the range of 10 Myr -- 15 Gyr with 5 possible values for the SFR within each bin, and we wished to test a small number of different IMF slopes, reddenings, and distance moduli, we would need to construct $>10^7$ model star formation histories.  As each model CMD needs to contain at least as many individual stars as the observed CMD, our available computing resources would be greatly overmatched.

A more convenient approach is to treat SFR($t$) as a free parameter for which we find the best solution for a given set of input parameters.  With this approach, we construct model CMDs representing a constant SFR in each age bin of the desired solution, constrained by the input Z($t$) relation, IMF, reddening, and distance modulus.  By solving for the coefficients of the linear combination of models that best fits the distribution of stars of the CMD, we recover SFR($t$).
The number of model star formation histories we need to construct is then vastly decreased, and is simply equal to the number of combinations of Z($t$) relations, IMF slopes, reddenings, and distance moduli we wish to test times the number of age bins in the solution.  An additional advantage of this approach is that a fit parameter can be used to test the goodness-of-fit of the models and errors can easily be calculated from Monte Carlo simulations.  An implementation of such an approach was first discussed by Dolphin (1997), who used $\chi^2$ as the fit parameter and singular value decomposition (Press et al.\ 1992) to solve for the best-fit star formation history from a suite of models with a number of different IMFs, reddenings, distance moduli, and abundances.  Dolphin applied the method to simulated CMDs to test the accuracy with which the input parameters could be recovered, but did not apply the method to actual observations.

We used the principles of this automated approach to construct a method to extract the star formation histories from our field star CMDs.  The method consists of four basic steps, which we describe in the following subsections.

\subsection{Interpolation of Isochrones}

\begin{figure}
\plotone{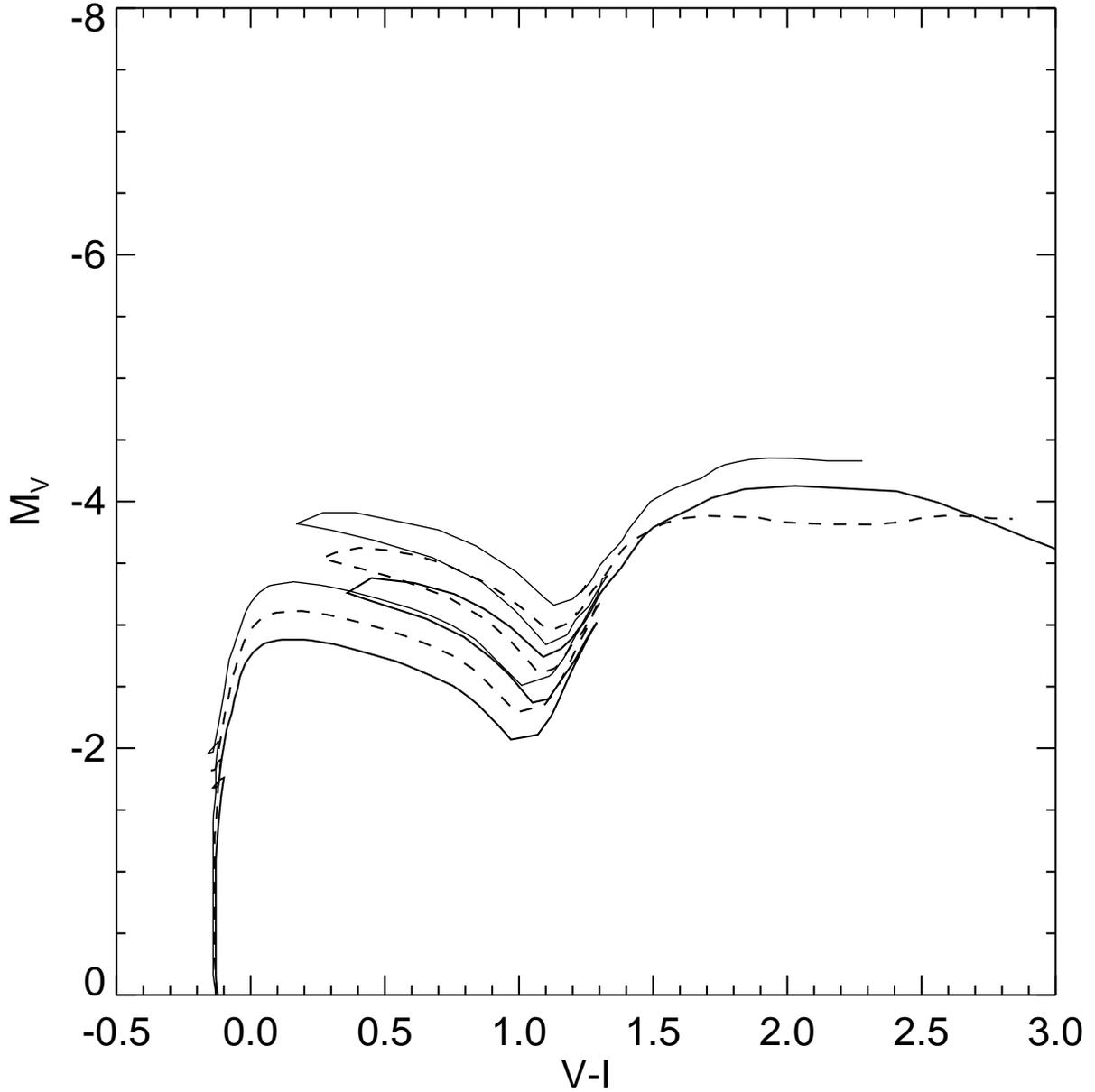}
\caption{An interpolated isochrone (solid lines) with $\log_{10}$ Age = 8.05 and [Fe/H]$\sim -$0.4 is shown with adjacent isochrones of $\log_{10}$ Age = 8.0 and 8.1 (dashed line).  While the interpolation faithfully reproduces the shape of the isochrone over the majority of evolutionary phases, it fails on the asymptotic giant branch.}
\label{probint}
\end{figure}

The B94 isochrones are well-suited for our purposes because they model every post-main sequence evolutionary phase, incorporate the recent OPAL opacities (Rogers \& Iglesias 1992), and are calculated for $V$ and $I$, our observed passbands.  Their main drawback is that they are difficult to interpolate, as they are not tabulated according to common evolutionary points.  To properly model the CMDs, we needed to interpolate the isochrones in metallicity as well as age.  We performed this interpolation by breaking the isochrones into segments based on the characteristic evolutionary phases listed in Tables 7$-$12 of B94, an electronic copy of which was kindly provided by G. Bertelli.  Each of these segments was then broken into a fixed number of mass bins.  A new isochrone could then be interpolated point-by-point from adjacent isochrones, with the segments ensuring that the isochrone shape was preserved.  We encountered a few problems, however.  A problem of a minor nature was that we discovered that the $V-I$ and $V$ for the masses of the evolutionary phases listed in B94's Tables 7$-$12 were often inconsistent with those tabulated in the isochrones.  Instead of using $V-I$ and $V$ from Tables 7$-$12, we therefore used the $V-I$ and $V$ from the isochrones for the points nearest the masses in the Tables.  Another small difficulty is that the number of evolutionary points listed in B94's Tables 7$-$12 occasionally changes for isochrones of different ages and different metallicities.  For isochrones affected by these changes, we used our best judgement in deciding the correct mapping.  A more serious problem is that large changes in the curvature of the isochrones do not always correspond with evolutionary points listed in the tables.  Fig.\ \ref{probint} illustrates this problem.  The heavy lines are isochrones with log ages of 8.0 and 8.1 years and [Fe/H]$\sim$-0.4, while the light line is an interpolated isochrone with log age of 8.05.  While the interpolated isochrone faithfully reproduces the shape of the adjacent ones over most of the isochrone, even on the blue loop, it deviates on the asymptotic giant branch.  This is because there is no evolutionary point in Table 10 of B94 designating the finish of the blue loop and the start of the AGB.  While we found that the interpolated isochrones have a number of areas where they deviate from the correct shape, the deviations occur in short-lived phases of stellar evolution.  We therefore did not attempt to correct the problems, as few of the stars in our observed CMDs will occupy these phases.

\subsection{Construction of the Model Star Formation Histories}

\begin{figure}
\plotone{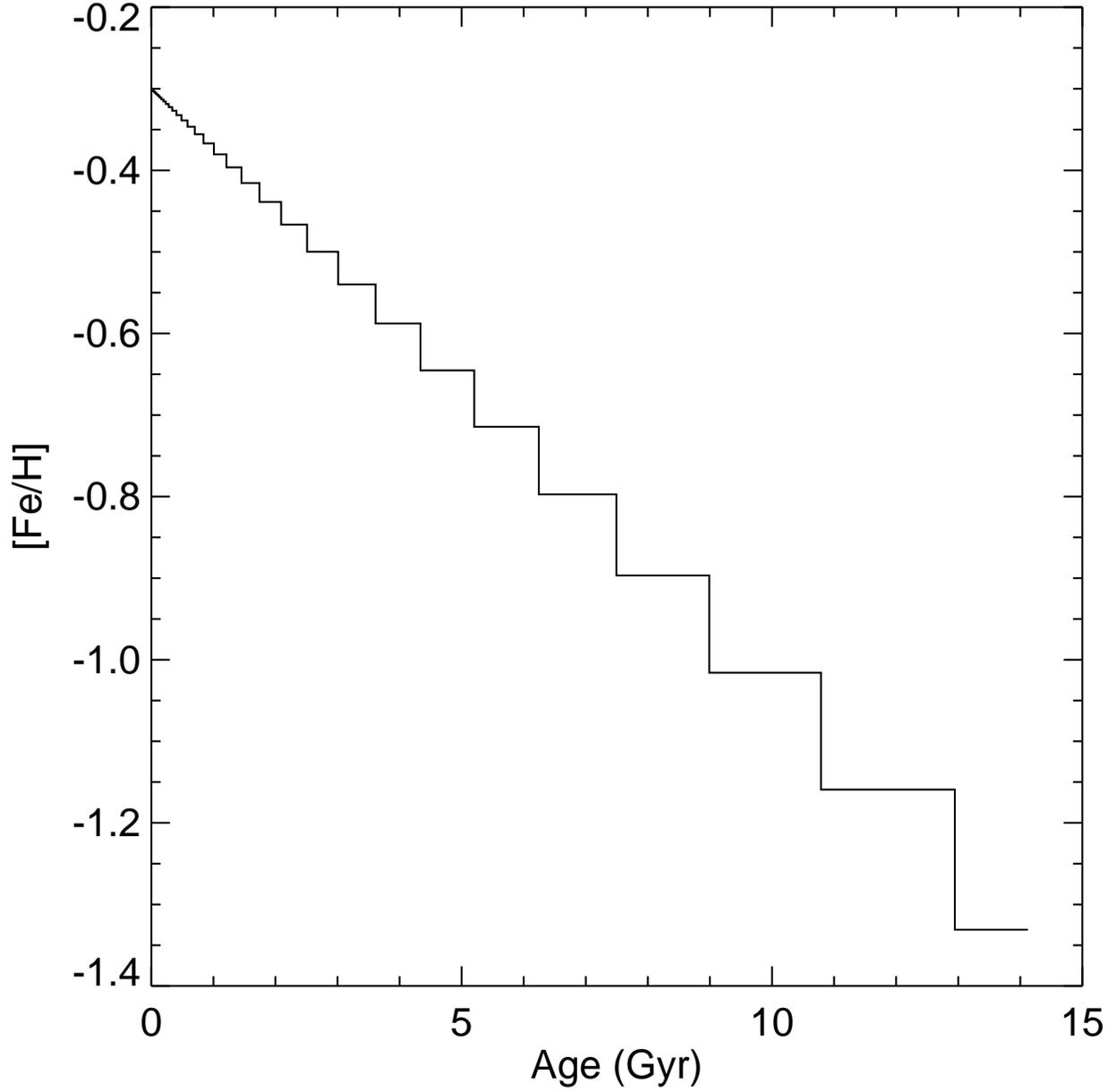}
\caption{A linear fit to the LMC cluster age-metallicity relation, adapted from Olszewski et al.\ (1996).  At the metal-poor end, the average abundance of the globular clusters studied in Olsen et al.\ (1998) has been used.}
\label{chemevol}
\end{figure}

We produced model CMDs using reddenings of $E(B-V)$=0.04$-$0.12 with steps of 0.02, distance moduli of 18.4$-$18.7 with steps of 0.1, and IMF slopes of x=1.5, 2.0, 2.35, and 3.0, where the IMF is given by the proportionality $N(m) \propto m^{-x}$.  The combinations of parameters are summarized in Table \ref{mpars}.  The ranges of $E(B-V)$ were chosen to span the likely range for these fields, the distance moduli include both the ``short'' and ``long'' distance scales of the LMC, and the IMFs were picked to span the range of a number of derivations of the IMF in the LMC.  The age-metallicity relation adopted is linear in age with a slope of -0.073 dex Gyr$^{-1}$ and a zero point of -0.3 dex.  This model, shown in Fig.\ \ref{chemevol}, agrees with the age-metallicity relation suggested by LMC clusters (Olszewski et al.\ 1996), with ours having a slightly higher metallicity at older ages that reflects our readjustment of the abundances of the inner globular clusters (Paper 1) over Olszewski et al.\ (1991).  However, because of the large gap in the LMC cluster age distribution, our adopted age-metallicity relation must be regarded as very uncertain over the $\sim$5$-$10 Gyr range.  Moreover, it is possible that the chemical evolution of the clusters and field stars is to some degree decoupled.  As discussed by Gallagher et al. (1996) in the context of the LMC, the degenerate effect of age and metallicity on stars in the CMD makes the distribution of metallicities a critical assumption, particularly for older stars.  Lack of knowledge about the field star metallicities thus necessarily weakens our final conclusions.

We constructed the model CMDs using 36 logarithmic age bins.
The bin widths, set to $\sim$0.08 in the log, and number of bins were chosen to approximately match the resolution with which we could hope to extract the star formation history from the observed CMDs.  We chose to use logarithmic age bins to compensate for the fact that isochrones are spaced more closely together at larger ages, decreasing the resolution with which one can discriminate between populations of different ages.
Within each age bin, we selected 20000$-$30000 stars with randomly distributed ages and masses, with limits on the ages imposed by the bin edges and limits on the masses set approximately equal to the maximum and minimum masses of stars on the isochrones within the age bin.  The number of stars was chosen to greatly exceed the number expected in the observed CMDs so as to limit counting statistics in the models, yet be smaller than the number of artificial stars available for simulating the errors.  In order to increase the sampling of stars with higher masses, we selected the masses from a flat IMF (x=0.0) but assigned each star a weight as if it were selected from the desired IMF of the model.  After all stars were selected, we calculated the total weighted mass in stars within each age bin, adding to this total the integral of the IMF over the range 0.08 $\le M \le$ 120 M$_\odot$ but excluding the mass range occupied by the simulated stars.  The total mass represented in the age bin was then used to calculate the star formation rate within each age bin.

For each model star, we linearly interpolated $V-I$ and $V$ from the nearest isochrones of appropriate age and metallicity.
Photometric errors were applied to these stars through the use of the artificial star tests described in Paper 1.  For each star in the simulated CMD, we randomly selected a nearby artificial star.  If the artificial star was not recovered during photometry, we removed the star from the simulated CMD.  Otherwise, the star was moved according to the photometric shifts experienced by the artificial star.  To increase the number of artificial stars from which we could simulate the errors, we allowed the program to pick artificial stars of a restricted range in magnitude but of any color.  Because stars experience systematic photometric shifts that depend on color, we corrected for the different systematic shifts before moving the model stars.

\begin{figure}
\plotone{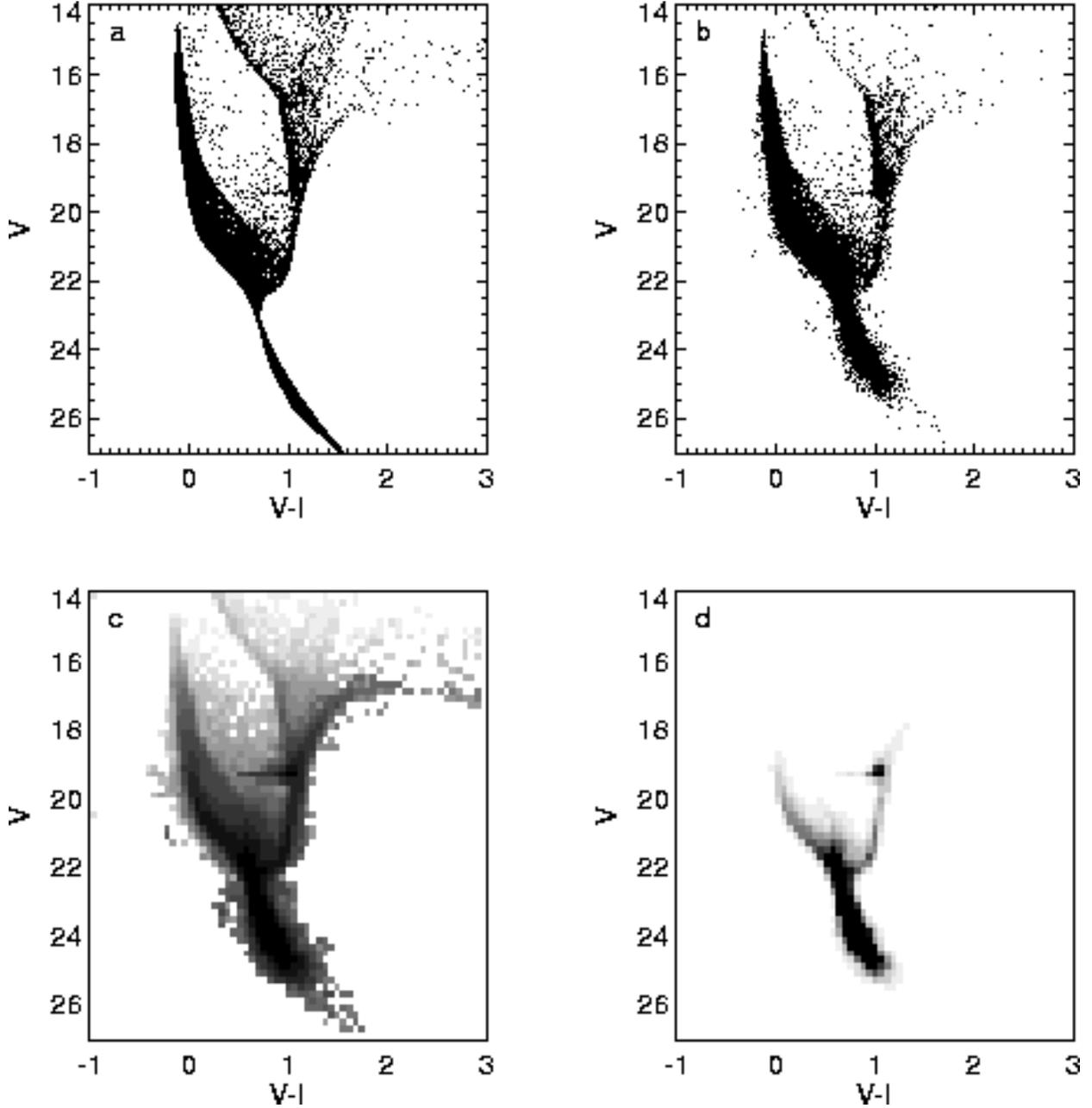}
\caption{$a)$ A sample model color-magnitude diagram, showing all 36 age bins of the model at once.  The stars have been selected from a flat IMF, constrained by the age-metallicity relation of Fig.\ \ref{chemevol}, and given an $E(B-V)$ of 0.08 and $(m-M)_\circ$ of 18.5. $b)$ The CMD shown in $a$ after application of photometric errors calculated from artificial star tests. $c)$ Hess diagram of $b$, scaled logarithmically to bring out the weaker features.  Scalings have been applied so that the diagram simulates a Salpeter IMF and a constant star formation rate of 1 M$_\odot$ yr$^{-1}$. $d)$  Same as $c$, scaled linearly.  The maximum greyscale is half that of $c$.}
\label{models}
\end{figure}

We stored the simulated CMDs as Hess diagrams with bins of 0.0625 in $V-I$ and 0.2 in $V$.  At this stage, the weights to bring the CMDs into keeping with the desired IMF were applied and the histograms were scaled so that each age bin represents a star formation rate of 1 M$_\odot$ yr$^{-1}$.  Figs.\ \ref{models}a-d show a sample raw simulated CMD, the simulated CMD with photometric errors, and the binned CMD, all summed over the 36 age bins.

\subsection{Comparison of Model CMDs with Observed CMDs}
For deriving SFR($t$) and identifying the best-fit model, the $\chi^2$ parameter tested by Dolphin (1997) is not ideal, as $\chi^2$ is highly sensitive to outlying points.  We instead chose to minimize the following parameter, which is derived from the Lorentzian distribution (Press et al.\ 1992):
\begin{equation}
P_{\rm Lor} = \sum_i \log\left(1 + \frac{1}{2}\left(\frac{O_i-M_i\left(\vec{x}\right)}{\sigma_i}\right)^{2}\right)
\end{equation}
where $O_i$ are the observed number of stars in the CMD at point $i$ in the $(V-I,V)$ grid, $M_i(\vec{x})$ is the number of stars predicted at grid point $i$ by the model with parameters $\vec{x}$, and $\sigma_i$ is the uncertainty in the number of stars at $i$, which we calculated from the models assuming Poisson statistics.
The Lorentzian, with its large tails, is much less sensitive to outliers than $\chi^2$.
We expect to find outliers when comparing our observations to the models for several reasons.  A primary reason is that the model isochrones used to generate the model CMDs do not perfectly describe the process of stellar evolution, especially for the evolved phases.  In addition, we are assuming that a single reddening, distance modulus, and IMF is appropriate for all of the stars in a single CMD, and that all stars of a single age have the same metallicity,
while in reality there will be some spread in these quantities.  Finally, it is unlikely that we have exactly reproduced the photometric errors through our artificial star tests, and there could be significant tails in the photometric distributions that we have not modeled correctly.
However, as long as the bulk of the stars are described by our single parameters and our models of the photometric errors apply to most of the stars, the remaining stars may be treated as outliers.  For the NGC 1916 field, the assumption of a single reddening is clearly strongly violated, so that differential reddening must be included for a viable model fit.  

\subsection{Method for Solution of SFR($t$)}
For each model with different $E(B-V)$, distance modulus, and IMF we can solve for SFR($t$) by minimizing equation 1 over all of the model age bins.  Because our models are normalized to represent a star formation rate of 1 M$_\odot$ yr$^{-1}$, the best-fit linear combination will produce SFR($t$) in units of M$_\odot$ yr$^{-1}$ over the area of the field.
A partial code for producing the solutions was kindly provided by A. Dolphin, which we rewrote and amended for our purposes.
The program minimizes $P_{\rm Lor}$ using the {\it Numerical Recipes} routine {\tt amoeba} (Press et al.\ 1992).  Given a starting guess at the correct values of the fitted parameters and the likely scale of the parameters, {\tt amoeba} searches the $N$-dimensional parameter space for a minimum in the surface, stopping when small changes in the parameters change the value of the function by less than a given amount, which we set to be slightly larger than the machine precision.  Because {\tt amoeba} stops when a minimum is found, the minimum may not be the desired global minimum.  To more thoroughly search the $P_{\rm Lor}$ surface for a global minimum, {\tt amoeba} was restarted several times with the new input guess and scale of the parameters set equal to the previous output.  After 4-5 restarts, {\tt amoeba} returned to the same minimum as the previous run, indicating that no other nearby minimum exists.  In this fashion, we established solutions of SFR($t$) for each model with different $E(B-V)$, distance modulus, and IMF.  For each IMF, the best-fit $E(B-V)$ and distance modulus were decided by picking the model with the lowest value of $P_{\rm Lor}$.

\section{Tests}
\begin{figure}
\plotone{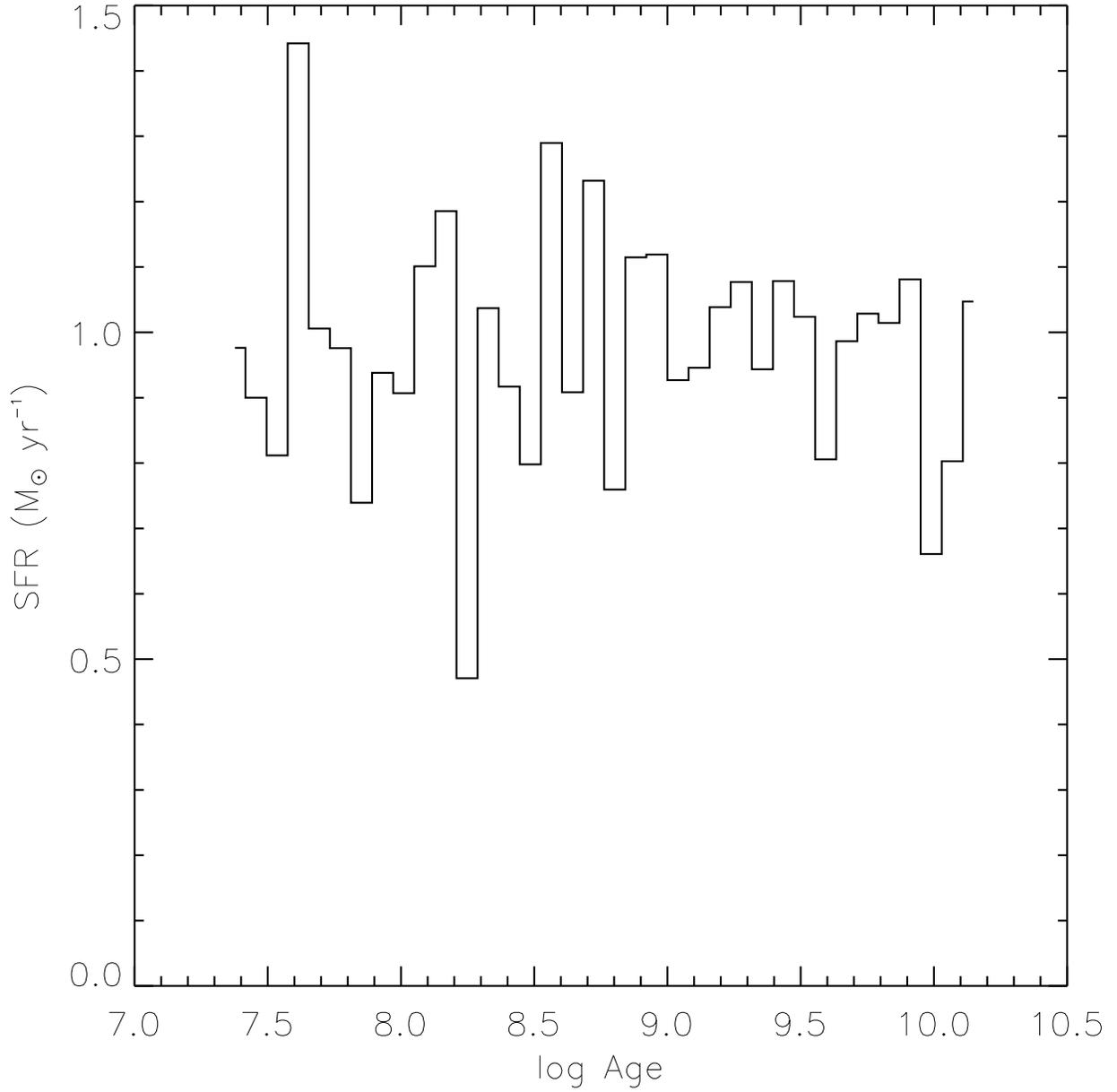}
\caption{Input star formation history used in tests of the solution method described in the text.}
\label{sfrin}
\end{figure}

Dolphin (1997) has shown, using a method very similar to the one described in this work, that it is possible to faithfully extract the star formation history from a simulated CMD representing a typical Local Group dwarf galaxy.  An outstanding question, however, is the age resolution with which the star formation history may be solved.  We explored this issue, as well as the ability of our method to solve for SFR($t$) and identify the correct $E(B-V)$ and $(m-M)_\circ$, by producing a CMD representing a simple star formation history and running it through the solution procedure.  The test case was constructed by taking a model with a uniform star formation rate and adjusting the star formation rate in each age bin by a randomly selected amount.  To avoid using pre-existing knowledge of the star formation history to our advantage during the solution, the input star formation history was kept secret until after the solutions were made.
Choosing $E(B-V)$ = 0.08, $(m-M)_\circ$ = 18.5, and x = 2.35 for the IMF slope, we simulated a CMD for the test case by selecting 1000 stars in each age bin from the interpolated B94 isochrones, to which we applied the completeness fractions and photometric uncertainties derived for the NGC 1754 WF frames, using the artificial star tests of Paper 1.  We binned this simulated CMD to the resolution of the models.  We also smoothed both the test CMD and the model CMDs with a Gaussian having a dispersion of one bin width, as we found that this limited the stochastic effects imposed by the binning.
Fig.\ \ref{sfrin} shows the input star formation history for the test case.

\begin{figure}
\plotone{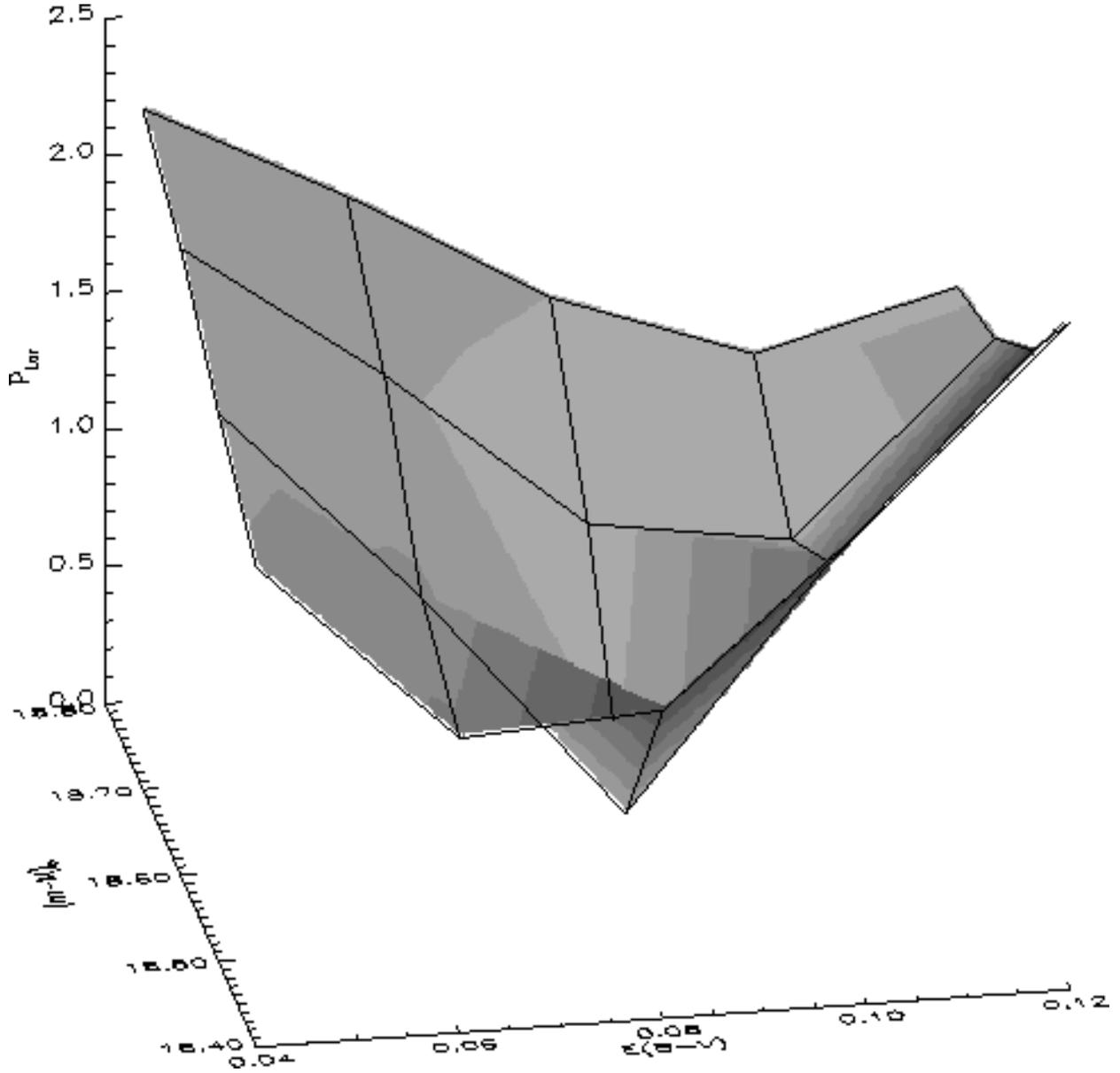}
\caption{P$_{\rm Lor}$ surface used to select the best-fit parameters of the solution to the star formation history in Fig.\ \ref{sfrin}.  The minimum occurs at $E(B-V)$=0.08 and $(m-M)_\circ$=18.5, which were the input values.}
\label{fitsurf}
\end{figure}

\begin{figure}
\plotone{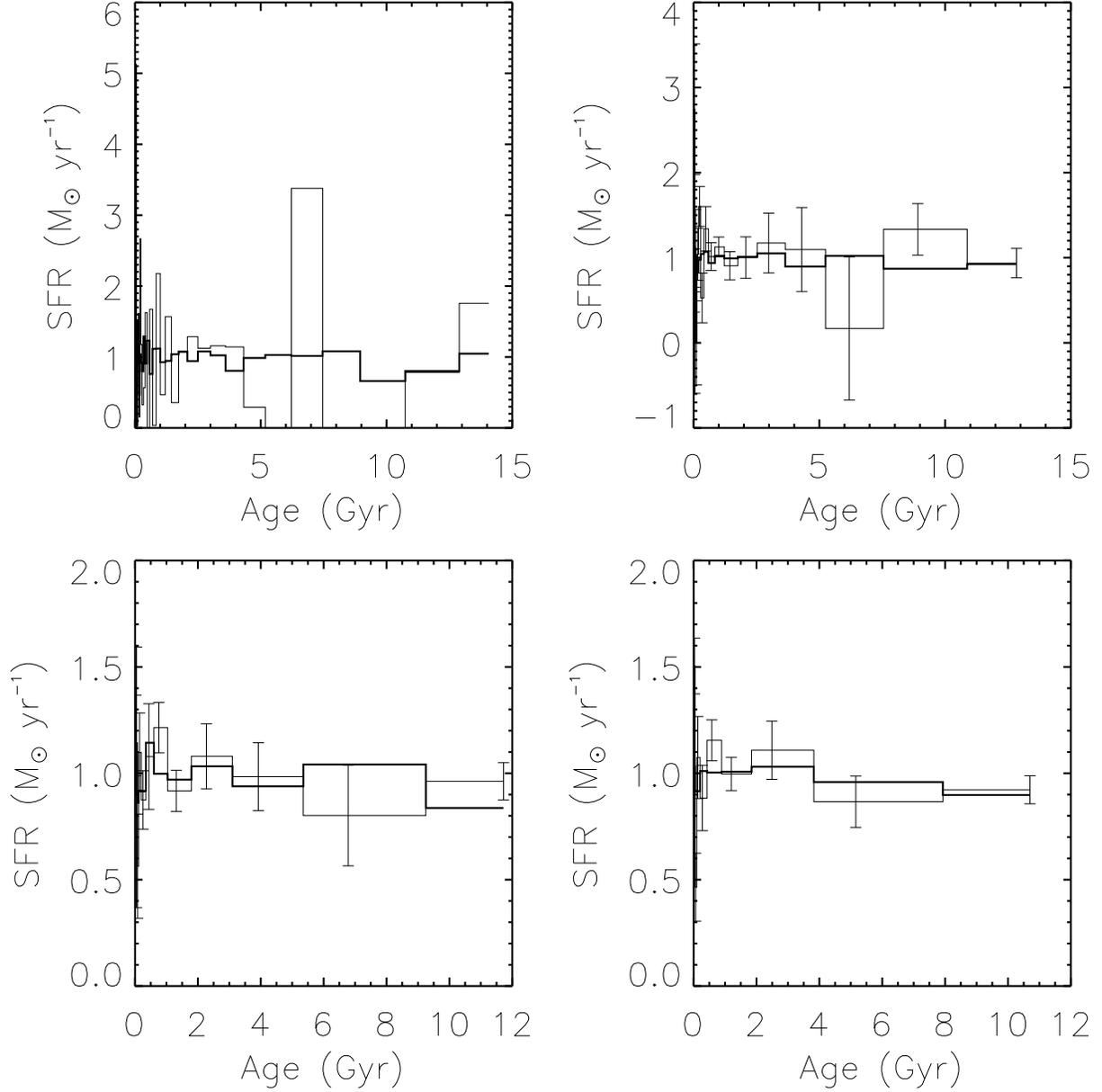}
\caption{$a)$ The thin line shows the star formation history derived from a color-magnitude diagram with stars selected from the profile shown by the thick line, using 36 logarithmic age bins in the solution.  The noisy solution indicates that the achievable age resolution is lower than that which was attempted. $b$, $c$, and $d)$ Solutions using 18, 12, and 9 age bins, showing the reduction in noise as the number of age bins is decreased.  Error bars have been calculated from repeated solutions obtained from re-sampled CMDs.}
\label{tests}
\end{figure}

Making a starting guess at the star formation rate in each bin, we used 4 iterations of {\tt amoeba} to find the best-fit coefficients of the models in each of the 36 age bins and each value of $E(B-V)$ in the range 0.04$-$0.12, $(m-M)_\circ$ in the range 18.4-18.7, and x = 2.35.  Fig.\ \ref{fitsurf} shows the computed $P_{\rm Lor}$ surface as a function of $E(B-V)$ and $(m-M)_\circ$.  The minimum of the surface occurs at $E(B-V)$ = 0.08 and $(m-M)_\circ$ = 18.5, in agreement with the input values.  Fig.\ \ref{tests}a shows the star formation history derived for the minimum $P_{\rm Lor}$, compared with the input.  The output contains wild disagreements with the input star formation history, mimicking a ``bursting''-type star formation history.  Guessing that the problem is one of age resolution, we redid the solutions after grouping several age bins of the models together.  Figs.\ \ref{tests}b-d show the best-fit solutions to the star formation history after grouping every 2, 3, and 4 models together.  These solutions reproduce much better the input model; the disagreements are all within the uncertainties calculated from repeated test runs using resampled CMDs, which we describe below.

\begin{figure}
\plotone{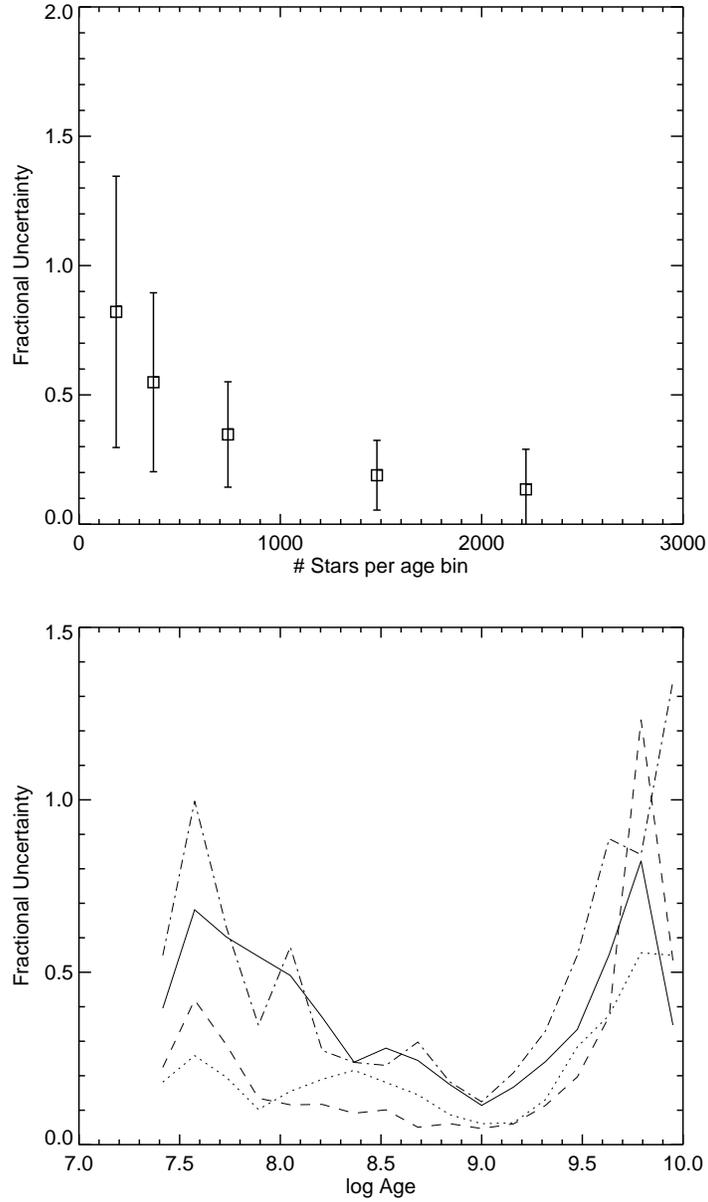}
\caption{{\it top panel)} Figure showing ${\rm \sigma(SFR_{in}-SFR_{out})/SFR_{in}}$ vs. number of stars per age bin for the test runs described in the text.  The error bars indicate the dispersion about the median $\sigma_{SFR}$, which was taken over all age bins of the solution.  {\it lower panel)} Figure showing ${\rm \sigma(SFR_{in}-SFR_{out})/SFR_{in}}$ vs. age, with a separate line for each run of tests.  The dashed, dotted, solid, and dashed-dotted lines represent the runs with 3000, 2000, 1000, and 500 stars per age bin, repectively.}
\label{resolution}
\end{figure}

Two factors contribute to the limiting age resolution of the star formation history solutions: the number of stars that fall within each age bin and the spacing of the models in the CMD.  The first factor is a function mainly of the total number of stars in the observed CMD while the second is a function of the age of the models and the photometric errors.  Grouping the models together has the effect of reducing the contribution of both factors; the number of stars within each age bin is increased while the spacing between the models also increases.  
To describe the effects that the number of stars and the model spacing have on the errors in the solution, we ran repeated tests using the model star formation history described above with varying numbers of stars in the simulated CMDs.  Table \ref{tpars} lists the parameters and number of runs of the tests.  For each set of runs, we calculated the dispersion in the recovered minus input star formation rates separately for each age bin of the solution.  The top panel of Fig.\ \ref{resolution} shows the dispersion in (${\rm SFR_{out}-SFR_{in}}$), taken as a median over all age bins in the solution, plotted against the number of stars per age bin in the CMDs used in each set of tests. 
The figure demonstrates that the median uncertainty in the star formation rates increases exponentially with decreasing numbers of stars per age bin, as is expected from Poisson statistics.  The lower panel of Fig.\ \ref{resolution} shows that the dispersion about the median over the age bins, indicated by the error bars in the top panel, is due to the effect of isochrone spacing.  At the youngest and oldest ages, the uncertainty in the
recovered star formation rate is higher than for intermediate ages, where model isochrones of similar ages are more easily distinguishable.  We will bear these effects in mind when discussing the age resolution of the star formation histories derived for the observed CMDs.

\section{Results for Field Star CMDs and Error Analysis}
\subsection{SFR($t$), Best-Fit Parameters, and Errors}
\begin{figure}
\plotone{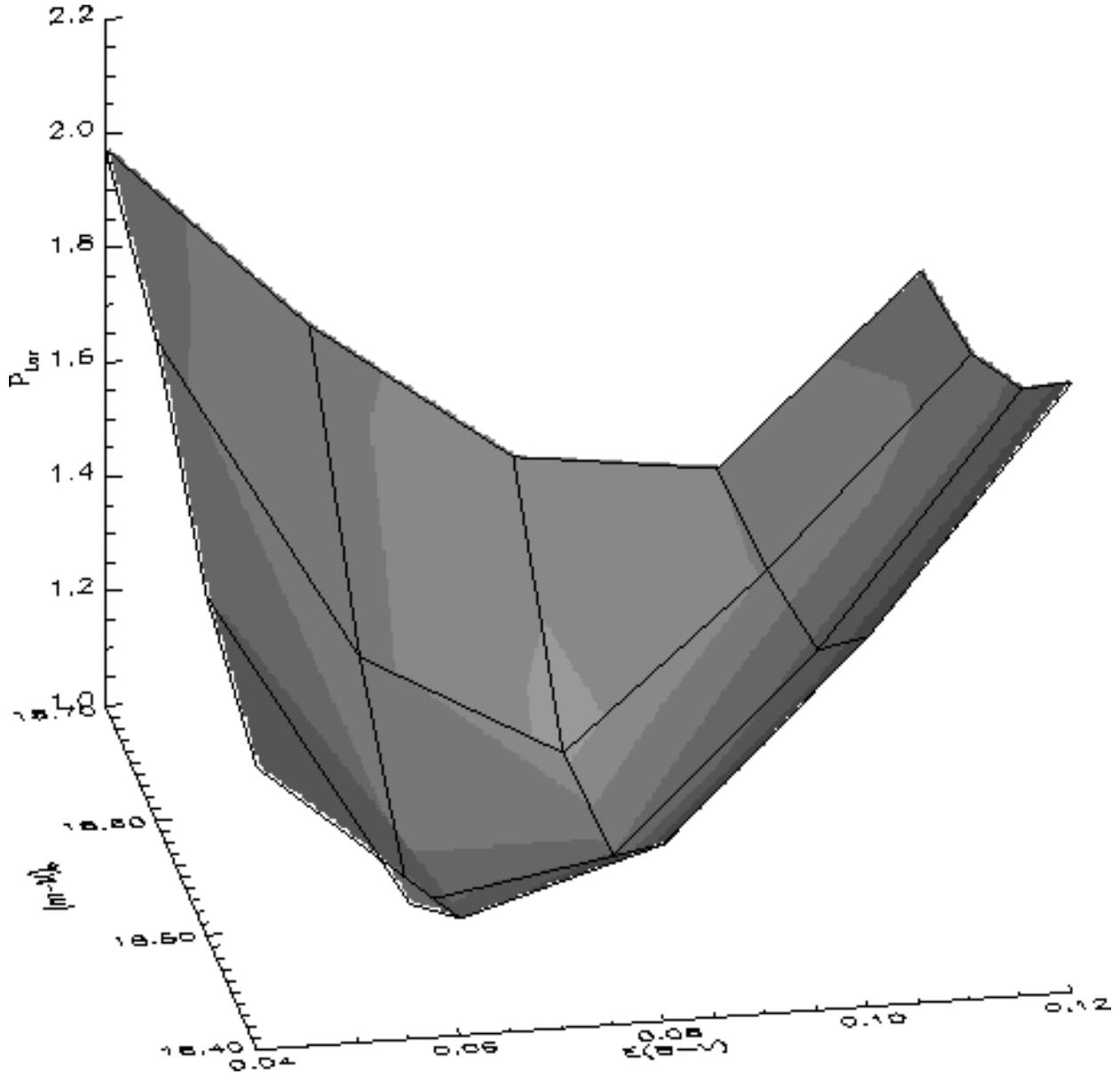}
\caption{P$_{\rm Lor}$ surface for solutions of the star formation history of the NGC 1754 field for the Salpeter IMF.  The minimum of the surface is at $E(B-V)$=0.06 and $(m-M)_\circ$=18.5.}
\label{n1754fitsurf}
\end{figure}

The SFR($t$) solutions for the LMC field star CMDs were made following the procedure outlined in section 4.  As was done for the tests described in section 5, we binned the observed CMDs to the same resolution as the models and smoothed both the observed and model color-magnitude grids with a Gaussian having a 1-pixel dispersion.  
We produced separate solutions for each combination of the parameters listed in Table \ref{mpars} and for 18, 12, and 9 age bins.  For NGC 1835 and NGC 2005, we found that we needed to add additional values of the reddening and distance modulus to bracket the full range of best-fit solutions.  For each IMF, we found the reddening and distance modulus that best match the observed CMD by choosing the minimum of the $P_{\rm Lor}$ surface, as was done for the test.  A sample surface is shown in Fig.\ \ref{n1754fitsurf}.  While the minimum $P_{\rm Lor}$ is well-defined, its value is not as low as those found during the tests, a sign that the model isochrones as input do not agree perfectly with the observations.

We calculated the errors in SFR($t$), $E(B-V)$, and $(m-M)_\circ$ through Monte Carlo simulations using bootstrapped samples of the observed CMDs.  We ran 25 simulations for each value of the IMF; with each run we resampled the observed CMDs and produced solutions of SFR($t$) for each value of $E(B-V)$ and $(m-M)_\circ$, exactly as was done for the analysis of the true observed CMDs.  Again, we chose the best-fit solution from the set by picking the minimum of the $P_{\rm Lor}$ surface.  

Figs.\ \ref{sfh}a-c show the derived star formation histories and associated errors using 18, 12, and 9 age bins in the solution and the Salpeter IMF, while Table \ref{ebv_dm} lists the best-fit values of 
$E(B-V)$ and $(m-M)_\circ$ for these solutions.
There are several common features in the derived star formation histories.  All of the fields show significant star formation in the last 3-4 Gyr with similar amplitude, with the NGC 1754 field, on account of its lower star density, proceeding at an overall rate $\sim$2$-$3 times lower than the rest.  The star formation has the appearance of a burst, peaking at $\sim$1 Gyr, in all of the fields.  While little can be said about the very young stars because of the small size of the fields and the close spacing of the young isochrones, there 
appears to be a decline in SFR($t$) over the last $\sim$0.5$-$1 Gyr.  This apparent recent decline could simply be evidence that the stars formed in the past 1 Gyr are not thoroughly mixed; the crossing time in the Bar is $\sim$80 Myr, implying a mixing time of a few hundred million years.  All of these features appear independent of the number of age bins used in the solution.

The NGC 1754 field star formation history appears to agree with that derived by H97 for a field near NGC 1866; both show an increase in the recent ($<$4 Gyr) star formation rate by a factor of $\sim$3, although the exact number depends on the age-resolution of the solution.  However, we find that the star formation did not instantly increase, but did so steadily.  We do not find evidence for the short burst of star formation occurring $\sim$2 Gyr ago that was found by Gallagher et al.\ (1996).

While the NGC 1754 field seems to only recently have formed a large number of stars, star formation in the Bar fields was active as long as $\sim$5$-$8 Gyr ago.  The existence of this older population in the Bar fields naturally explains the differences in RGB and SGB morphology seen in the CMDs, as the Bar fields clearly have broader RGBs and SGBs than the NGC 1754 field.  The older population is most clearly seen in the solution containing 9 age bins, but remains in the solutions with higher age resolution.    
This result implies that there is a significant difference in the star formation histories between the inner and outer regions of the LMC.  Because a bar does not confine any given star to the bar pattern, we of course cannot directly age-date the LMC Bar through its stellar populations.  However, the older population of stars that we see within the Bar could be readily explained by a bar instability triggering star formation as early as 5$-$8 Gyr ago.
This interpretation disagrees with that of Elson, Gilmore, \& Santiago (1997) who have suggested, on the basis of a WFPC2 CMD of a field in the Bar, that the Bar formed $\sim$1 Gyr ago, or 1$-$2 Gyr after the LMC disk.  Our results agree with Elson et al.\ in that we do find a peak in the star formation rate occurring $\sim$1 Gyr ago in the Bar fields; however, we also see this peak in the disk field.  Our interpretation also disagrees with the suggestion of Ardeberg et al.\ (1997) that the Bar may be younger than 0.5 Gyr.  Again noting that 0.5 Gyr is of order the mixing time in the Bar, it is quite possible that the large young population seen by Ardeberg et al.\ in their field represents a local star-forming event.
Our results also do not fully agree with those of Hardy et al.\ (1984), who found that no significant star formation has occurred in the Bar before 3 Gyr ago.  However, Hardy et al.\ may not have been able to detect an older population, as their CMD was not deep enough to reach the TO and lower MS.

An encouraging feature of each of the star formation histories of Figs.\ \ref{sfh}a-c is a significant number of stars in the last age bin, where we expect the contaminating globular cluster stars to be found.
How much of the old stellar populations can be attributed to the clusters spilling into the WF frames?  While we found we could not cleanly statistically subtract the cluster stars from the field star CMDs, we can use the estimates of their numbers derived previously to address the question.  In Table \ref{contam} we compare the number of cluster stars in the WF frames estimated from the King model fits to the number of stars in the last age bin of the star formation histories of Fig.\ \ref{sfh}a.  We find that the cluster stars contribute significantly to each field, while in the NGC 1754 and NGC 1898 fields they dominate the oldest bin.  This finding leads us to suppose that the narrow RGB of the NGC 1754 field is produced almost entirely by cluster contamination, while the broad RGBs and SGBs of the Bar fields truly represent an old field population.

\subsection{Goodness of Fit of the Model Star Formation Histories}
So far, we have implicitly assumed that the best-fit model star formation histories provide good fits to the observations.  However, our models make a number of simplifications over what is realistically expected; e.g. we assumed uniform reddening, no binary stars, and a simple linear age-metallicity relation.  To get an idea of whether the derived star formation histories provide sensible matches to the data, we overlaid isochrones representing peaks in the derived star formation histories on the field star CMDs.  Fig.\ \ref{isochrones} shows the ages at which the isochrones were chosen and the corresponding matches with the CMDs.  The inability to tell by eye whether the chosen peaks in the star formation rate correspond to increases in the stellar density in the CMD reinforce the need for an automated technique to derive the star formation histories.  We can tell, however, that there is a clear difference between the NGC 1754 CMD and the CMDs of the Bar fields.  While the NGC 1754 field appears to have a single old turnoff, the Bar fields have many stars $\sim$0.5 magnitude brighter than the oldest turnoff, which we interpret as the 4-8 Gyr population in Figs.\ \ref{sfh}a-c.  Overlaying the isochrones also shows that our adopted age-metallicity relation does not provide an ideal match to the CMDs.  The most striking disagreement is that the red giant branches of the older isochrones fall too far to the red to match the observed giant branch.  While the observed giant branch could be fit by older isochrones if the distance modulus were decreased, it would then be difficult to match the blue envelope of the upper main sequence, which is reproduced well with the current models.  A possible explanation is that the metallicity of the model old populations is too metal rich.  Because the metallicity of the oldest age bin of the model was chosen to equal the average abundance of the old globular clusters (Paper 1), this would imply that the field contains an old population that is considerably more metal-poor than the clusters.  However, because the globular cluster sequences fall quite closely on top of the field star RGBs, this would mean that the measured globular cluster abundances are too metal-rich by $\sim$0.5 dex.  As the globular cluster abundances were measured from the slopes of the RGBs, which are empirically calibrated with Milky Way clusters using high-dispersion spectroscopic abundances, we would be surprised if they were found to be so grossly incorrect.  Another possibility, however, is that the B94 isochrones do not correctly predict the color of the red giant branch for low metallicities; our comparison suggests that the model RGBs are too red by $\sim$0.08 magnitudes, given that we have solved for the correct reddening and distance modulus of the field stars.  

\begin{figure}
\plotone{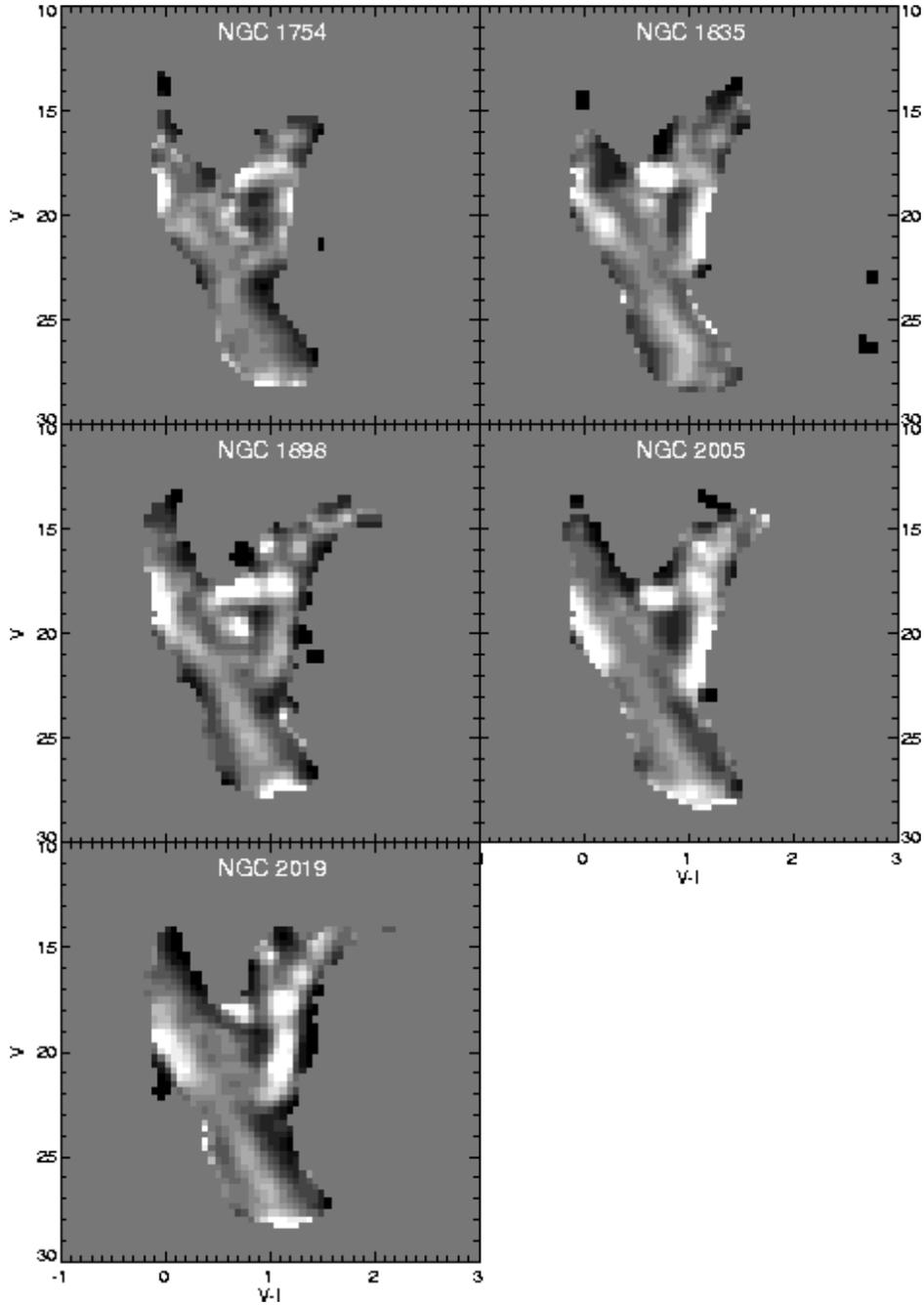}
\caption{Residuals of the best-fit model star formation histories of the {\it HST} fields and the observed CMDs, adopting a Salpeter IMF.  The residuals have been divided by the number of observed stars to remove the effect of the luminosity function.  Bright patches indicate an excess of predicted stars while dark patches are regions where there are more observed stars than predicted.  Grey areas indicate good agreement between the models and observations.  The grey levels have been stretched from -1 to 1.}
\label{residuals}
\end{figure}

\begin{figure}
\plotone{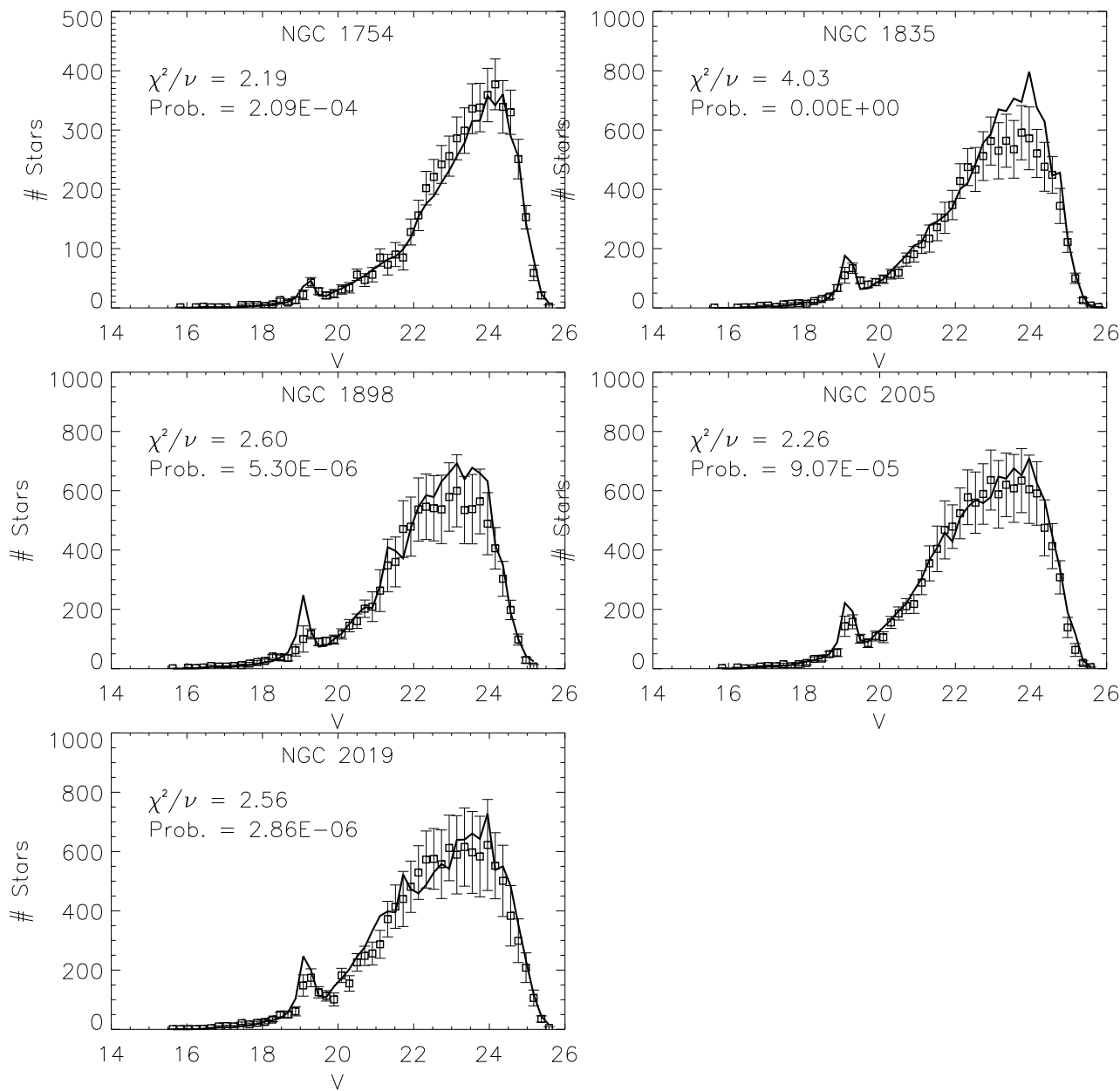}
\caption{Comparison of the observed luminosity functions (squares) with the best-fit model star formation histories (solid lines), adopting a Salpeter IMF.  The error bars on the points are those due to Poisson number statistics and the uncertainties in the derived star formation rates.  The figures indicate the $\chi^2$ per degree of freedom for the fits and the probability that $\chi^2$ would exceed the given values by chance.}
\label{lf}
\end{figure}

In Fig.\ \ref{residuals} we show the residuals produced by the models of Fig.\ \ref{sfh}a.  The residuals are shown as fractions of the observed star density in the CMDs.  The greyscale stretches from -1 to 1; white pixels are regions where too many stars are predicted by the models, black pixels represent an excess of observed stars, and grey pixels represent the regions where the models closely match the observations.  The residuals show that over large areas of the CMDs, the models
reproduce the observations fairly well.  However, as expected from the comparisons with isochrones, there are several areas where the fits are poor, most notably the area of the red giant branch, the horizontal branch and red clump stars, and a small section of the upper main sequence.
These problems are not as apparent if the model and observed luminosity functions are compared instead of the CMDs.  These luminosity functions are shown in Fig.\ \ref{lf}; the errors in the points include both Poisson counting statistics and the uncertainties in the derived star formation rates of the models.  The use of luminosity functions smooths over many of the disagreements between the models and observations; application of the $\chi^2$ statistic shows that the models are not strongly excluded, except for the case of the NGC 1835 field.

\subsection{Exploration of Systematic Errors on the Derived Star Formation Histories}
\begin{figure}
\plotone{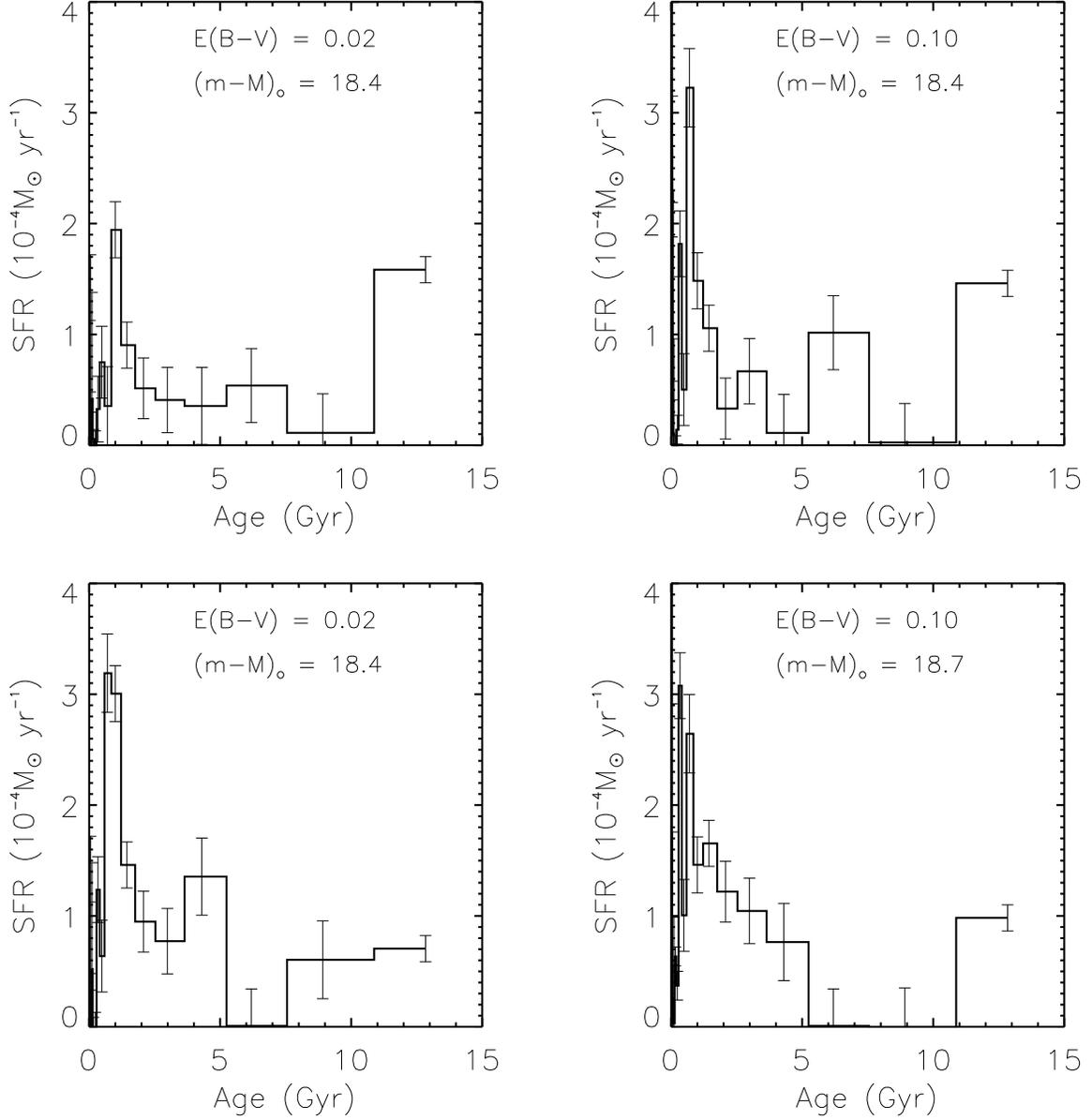}
\caption{$a)$In these four plots, $E(B-V)$ and $(m-M)_\circ$ have been set to extreme values and the effect on the NGC 1835 field star formation history shown.  In general, decreasing $E(B-V)$ or $(m-M)_\circ$ increases the contribution of older stars, while increasing these parameters has the opposite effect. $b)$Comparison of selected model isochrones with the NGC 1835 field CMD for the combinations of $E(B-V)$ and $(m-M)_\circ$ shown in $a$.}
\label{extremes}
\end{figure}

\begin{figure}
\plotone{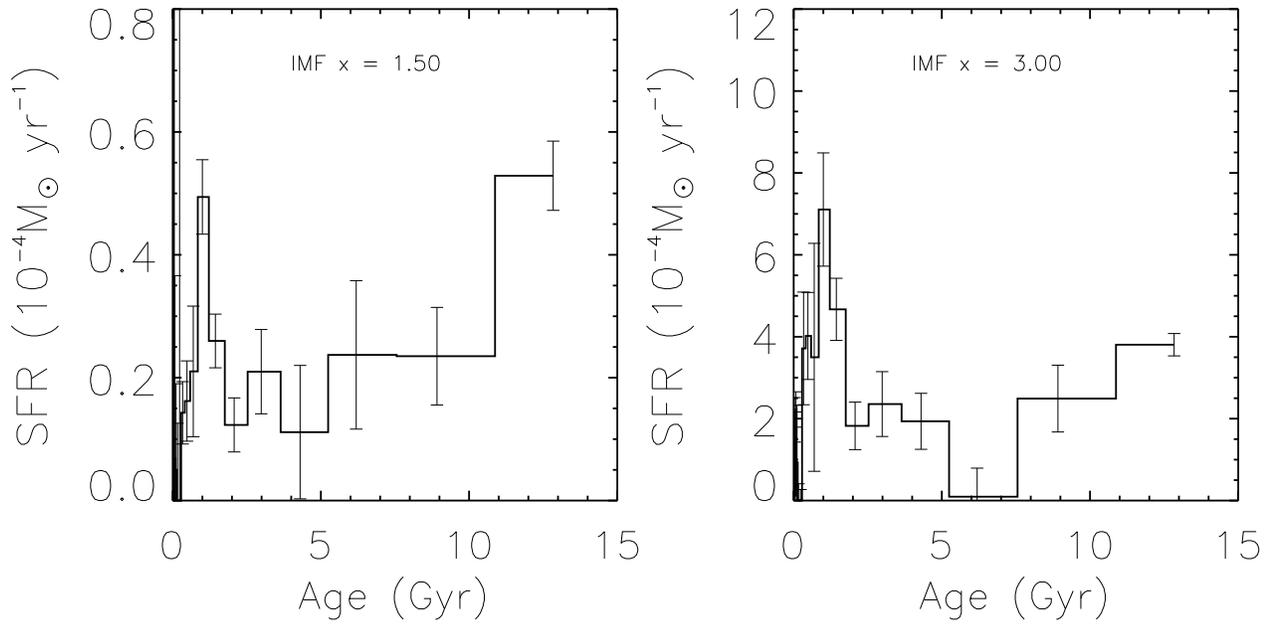}
\caption{Plots showing the effect of setting the slope of the IMF, $x$, to extremes on the star formation history of the NGC 1835 field.  Lowering $x$ increases the relative contribution of the old populations, while raising $x$ emphasizes the young stars.}
\label{imfvar}
\end{figure}

The star formation histories derived in Section 5.1 may be incorrect in detail because of the inconsistent match of the models and observations.  To explore the effects that the systematic disagreements between the models and observations might have on the star formation histories, we set the reddening and distance modulus to extreme values and examined the resulting star formation histories.  Fig.\ \ref{extremes}a shows the star formation history of the NGC 1835 field for the case of 18 age bins in the solution after setting $E(B-V)$ and $(m-M)_\circ$ to extremes.  In general, decreasing either $E(B-V)$ or $(m-M)_\circ$ increases the relative contribution of the old stellar populations, while increasing either one has the opposite effect.  This can be understood by Fig.\ \ref{extremes}b, where we plot isochrones having the extreme values of $E(B-V)$ and $(m-M)_\circ$ over the CMD.  For the low values of $E(B-V)$ and $(m-M)_\circ$, the old isochrones cover the lower main sequence and RGB while the young isochrones do not fit the main sequence as well.  For the high values, the old isochrones do not fit the turnoff and RGB, but the young isochrones follow the main sequence envelope.
We note that decreasing $x$, the slope of the IMF, has the same effect on the star formation history as decreasing $E(B-V)$ or $(m-M)_\circ$ (see Fig.\ \ref{imfvar}).  This is because with a shallower model IMF, more evolved stars are needed to reproduce the observed luminosity function (see also discussion in H97).

\begin{figure}
\plotone{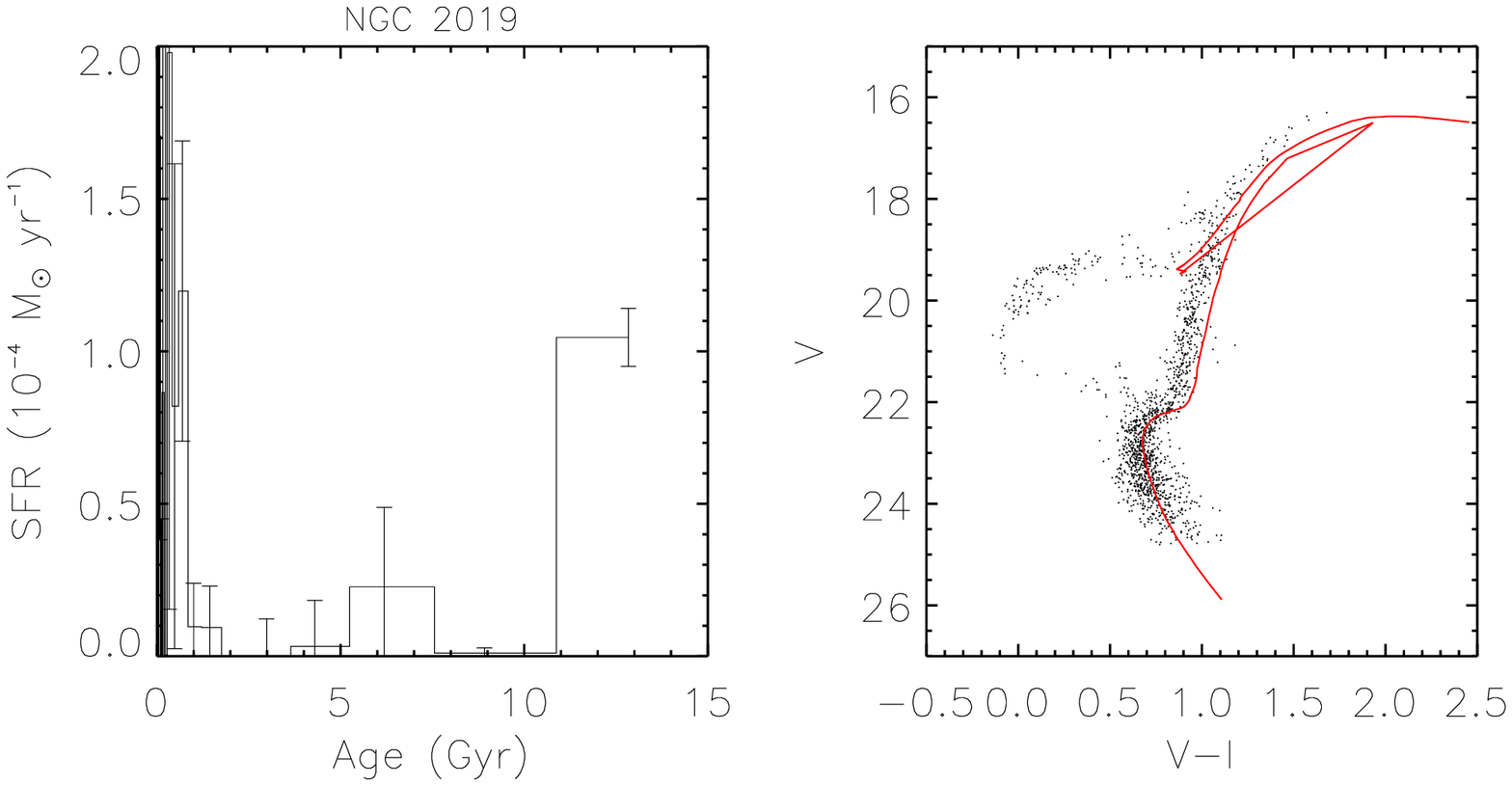}
\caption{Star formation history derived by using the method employed for the field star CMDs on the CMD of the NGC 2019 cluster.  As expected, the star formation history is dominated by the oldest age bin, while most of the remaining age bins have star formation rates consistent with zero within the uncertainties.}
\label{pcisoc}
\end{figure}

Thus, we might expect that the effect of the model disagreements would be to suppress the old populations in the derived star formation histories.
Because the disagreements are largest for the older stars, we should be able to quantify this expectation by using our method to solve for the star formation history of one of the globular cluster CMDs shown in Fig.\ 12 of Paper 1.  By using a cluster CMD with field stars statistically removed, we expect to recover a star formation history dominated almost entirely by the oldest bin.  In the left panel of Fig.\ \ref{pcisoc}, we show the star formation history derived for the NGC 2019 field star-subtracted cluster CMD.  We have adopted a reddening of $E(B-V)$=0.06 for the cluster, as derived in Paper 1, and a distance modulus of 18.5.  Indeed, the star formation history is characterized by a large peak in the oldest age bin with almost all of the remaining bins consistent with a star formation rate of zero within the errors.
In the right panel of Fig.\ \ref{pcisoc}, we show the model isochrone having age and metallicity equal to the average of the oldest bin of the solution.  While the isochrone fits the lower main sequence and turnoff luminosity, the model RGB falls too far to the red, in a fashion similar to the fits shown in Figs.\ \ref{isochrones}, and the HB morphology is not reproduced.  We find that the integral of the model luminosity function over the range 17 $\le V \le$ 23 is $\sim$50\% lower than that of the observed LF.  This result agrees with our expectation that the star formation rates in the oldest age bins of Figs.\ \ref{sfh}a$-$c have been underestimated due to the model disagreements.  Furthermore, because we do not see significant star formation in the age range 4$-$8 Gyr in Fig.\ \ref{pcisoc}, it appears unlikely that the model disagreements could be the cause of the different star formation histories derived for the NGC 1754 and Bar fields.


\section{Summary and Conclusions}
Forty years ago, Martin Schwarzschild remarked on the hopelessness of disentangling a color-magnitude diagram of stars that contains a wide mix of ages and compositions.  Modern computational power, aided by improved astronomical facilities, no longer makes the problem unapproachable.  In this work, we have applied an automated technique to the task of deriving star formation histories of five fields in the LMC from deep {\it HST} CMDs.  
However, assumptions still need to be made to make the problem tractable.  We have assumed throughout that the field star CMDs can be parametrized by a single reddening, distance modulus, and IMF; that binary stars may be ignored; that the IMF does not vary in time; and that a simple linear age-metallicity relation, based on LMC clusters, describes the enrichment history.

\begin{figure}
\plotone{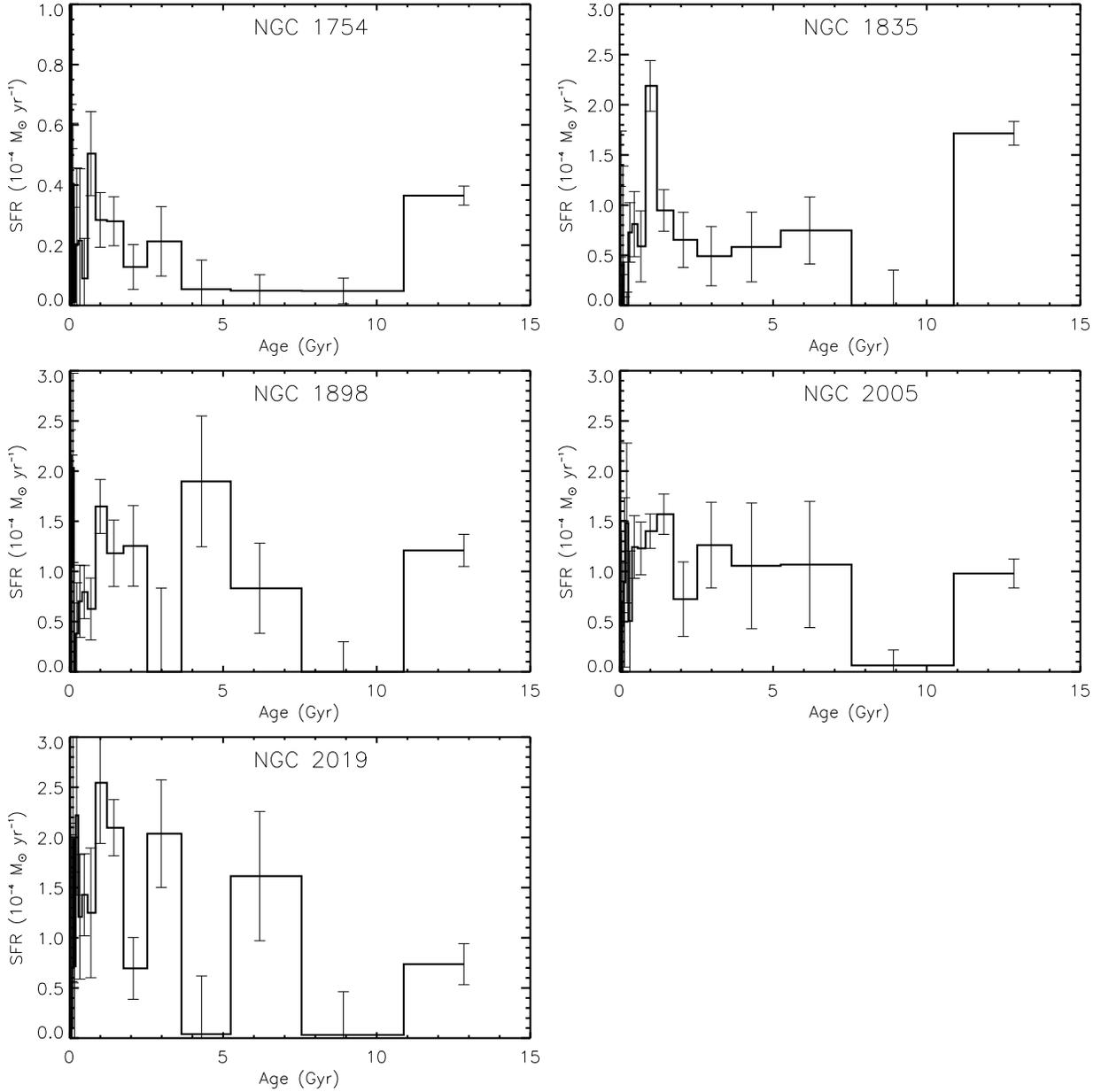}
\caption{$a)$ Star formation histories derived for the {\it HST} fields using the method described in the text, adopting 18 age bins in the solution.  In each field, the oldest age bin contains a contribution from the nearby globular cluster.  The error bars were calculated from Monte Carlo simulations using bootstrapped samples of the observed CMDs.  The NGC 1754 field star formation rates are shown on a scale 1/3 that of the other fields to bring out the detail in its star formation history. $b)$ Star formation histories derived adopting 12 age bins in the solution. $c)$ Star formation histories derived adopting 9 age bins in the solution.  While the loss in resolution over $a$ is significant, most of the salient features remain, and the error bars are smaller.}
\label{sfh}
\end{figure}

\begin{figure}
\plotone{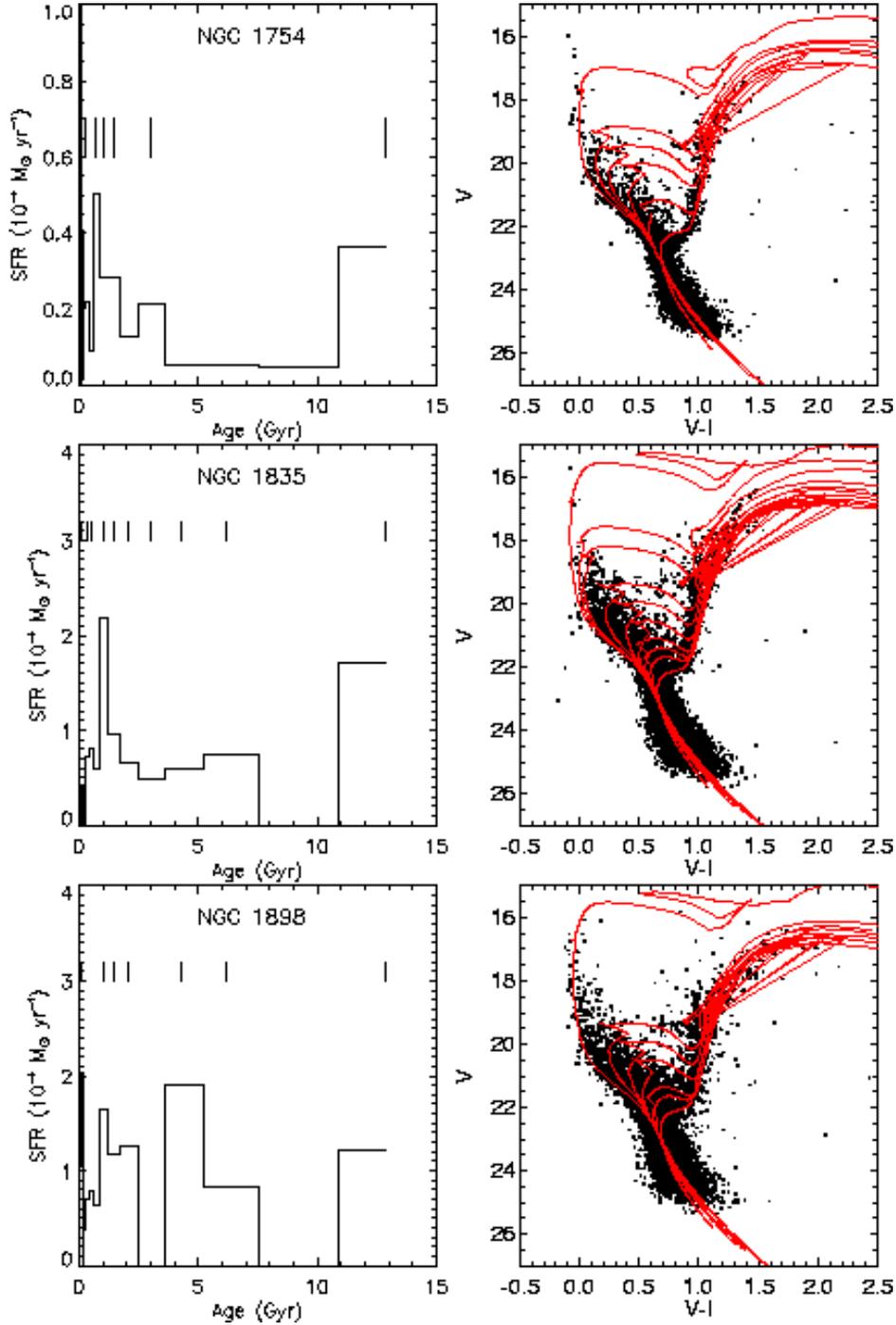}
\caption{In the left-hand column of plots, peaks and features in the star formation histories of Fig.\ \ref{sfh}a are marked by vertical lines.  Isochrones corresponding to the correct ages and metallicities of these features are overlaid on the CMDs in the right-hand column, using the derived reddenings and distance moduli.  While the model isochrones reproduce well the envelope of the upper main sequence, they fall too far to the red of the red giant branches.}
\label{isochrones}
\end{figure}

The star formation histories derived for the fields for a Salpeter IMF (Figs.\ \ref{sfh}a-c) contain a number of features that match our common-sense expectations.  All of the fields have a significant number of old stars confined to the last age bin.  As these fields lie within 1\arcmin \ of five old globular clusters, contamination from globular cluster stars is expected; in two of the fields, the contamination can account for the entire population of oldest stars detected in the field.

All of the fields show significant recent ($\lesssim3-4$ Gyr) star formation, in agreement with a large number of other LMC field star studies and with the LMC cluster age distribution.  In particular, the star formation history we derive for the NGC 1754 field agrees well with that derived by H97 for a field near NGC 1866, once the contribution due to contaminating globular cluster stars is ignored.  The decline in the star formation rate between the ages of 0.5 and 1 Gyr seen in each of the fields is naturally explained by inadequate mixing of stars formed most recently; the crossing time in the Bar is $\sim$80 Myr, implying a mixing time of a few hundred million years.

We find that the four Bar fields (NGC 1835, NGC 1898, NGC 2005, and NGC 2019) experienced significantly more star formation in the age range 4$-$8 Gyr than did the NGC 1754 field, which lies in the disk; this is most clearly seen in the solution containing 9 age bins (Fig.\ \ref{sfh}c).  The existence of this population of stars unique to the Bar naturally explains the broader giant branches and the existence of the stars 0.5 magnitudes brighter than the turnoff in each of the Bar fields but not in the NGC 1754 field.
This result implies that the LMC's inner region is older than has been suggested by e.g. Elson et al.\ (1997) and Ardeberg et al.\ (1997).

However, we also find disagreements between the best-fit models and observation; the biggest disagreement is in the region of the RGB, where the model sequences fall $\sim$0.08 magnitudes to the red of the observed sequence.  To explore the consequences of this disagreement, we set the reddenings and distances of the models to extreme values and examined the resulting star formation histories.  In addition, we examined the success of our method in deriving the star formation rate of the NGC 2019 cluster CMD.  Both explorations suggest that the model disagreements cause us to underestimate the contribution of the old populations.  Thus, we expect that the star formation histories of Figs.\ \ref{sfh}a$-$c are roughly correct {\it within} the assumptions made about the IMF, the effect of binary stars, and the age-metallicity relation.  Of these assumptions, the age-metallicity relation is on the weakest footing because of poor knowledge of LMC metallicities in the $\sim$5$-$10 Gyr range.

While we think the method used here to derive star formation histories has potential, it can clearly be considerably improved.  Helpful next steps would be to include binary stars in the models, explore the effect of different models of chemical evolution, and to test additional sets of isochrones.  Future analyses will hopefully benefit from future improved knowledge of LMC metallicities and from better understanding of the later phases of stellar evolution.  

\acknowledgements
I would like to thank Paul Hodge for his guidance throughout this project.  I also thank Bob Schommer, Nick Suntzeff, Alistair Walker, and Chris Smith for helpful discussions and for thorough readings of the original manuscript.  I thank the anonymous referee for several excellent comments which improved the manuscript.  Andy Dolphin provided the routines around which I built the program to extract the star formation histories.  This work was supported by STScI grant GO05916 to Nick Suntzeff.

\clearpage
\begin{deluxetable}{lccccccccc}
\small
\tablewidth{0pt}
\tablecaption{NGC 1754 Wide Field Camera Photometry}
\tablehead{
\colhead{Star \#} &
\colhead{X ($V$)} &
\colhead{Y} & 
\colhead{$V$} &
\colhead{$\sigma_V$\tablenotemark{a}} & 
\colhead{$I$} & 
\colhead{$\sigma_I$\tablenotemark{a}} & 
\colhead{$V$ type\tablenotemark{b}} & 
\colhead{$I$ type\tablenotemark{b}} & 
\colhead{Removed\tablenotemark{c}} \\
}
\startdata
0 & 214.82 & 76.38 & 20.2346 & 0.0330 & 20.1296 & 0.0390 & 1 & 1 & \\
1 & 184.19 & 81.18 & 19.5396 & 0.0360 & 19.4586 & 0.0330 & 1 & 1 & \\
2 & 273.77 & 82.37 & 20.2416 & 0.0300 & 20.1716 & 0.0340 & 1 & 1 & \\
3 & 96.77 & 91.67 & 19.3856 & 0.0240 & 18.4896 & 0.0250 & 1 & 1 & \\
4 & 506.60 & 105.65 & 19.3696 & 0.0300 & 18.2548 & 0.0260 & 1 & 10 & \\
5 & 513.20 & 109.20 & 19.8426 & 0.0360 & 18.8726 & 0.0310 & 1 & 1 & y\\
6 & 296.19 & 110.51 & 20.0626 & 0.0270 & 19.6656 & 0.0250 & 1 & 1 & \\
7 & 119.41 & 118.82 & 19.1496 & 0.0250 & 19.1116 & 0.0220 & 1 & 1 & \\
8 & 247.40 & 127.18 & 19.6766 & 0.0350 & 19.2936 & 0.0290 & 1 & 1 & \\
9 & 238.33 & 132.13 & 19.5236 & 0.0310 & 18.6766 & 0.0260 & 1 & 1 &  y\\
\dotfill & & & & & & & & & \\
\enddata
\tablenotetext{a}{Photometric errors reported by DoPHOT}
\tablenotetext{b}{Where the short exposure photometry was used, the DoPHOT object type has been multiplied by 10.}
\tablenotetext{c}{Stars removed by cluster star cleaning procedure are marked with ``y".}
\label{wfphot}
\end{deluxetable}

\clearpage
\begin{deluxetable}{lccc}
\small
\tablewidth{0pt}
\tablecaption{Expected Cluster Contamination}
\tablehead{
\colhead{Field} &
\colhead{$N$(cluster stars)} &
\colhead{$\frac{N({\rm cluster ~stars})}{N({\rm total})}$} & 
\colhead{$\frac{N({\rm oldest ~bin})}{N({\rm total})}$}\\
}
\startdata
NGC 1754 & 2031 & 0.40 & 0.51 \\
NGC 1835 & 708 & 0.07 & 0.48 \\
NGC 1898 & 2353 & 0.23 & 0.29 \\
NGC 2005 & 1130 & 0.10 & 0.25 \\
NGC 2019 & 1000 & 0.08 & 0.18 \\
\enddata
\label{contam}
\end{deluxetable}

\clearpage
\begin{deluxetable}{cccr}
\scriptsize
\tablewidth{0pt}
\tablecaption{Parameters of Model Star Formation Histories}
\tablehead{
\colhead{$E(B-V)$} & 
\colhead{$(m-M)_\circ$} &
\colhead{IMF $x$} &
\colhead{Fields Modelled} \\
}
\startdata
0.02 & 18.4-18.7 & 1.5 & NGC 1835 \\
0.04 & 18.3 & 1.5 & NGC 2005 \\
0.04 & 18.4-18.7 & 1.5 & all \\
0.06 & 18.3 & 1.5 & NGC 2005 \\
0.06 & 18.4-18.7 & 1.5 & all \\
0.08 & 18.3 & 1.5 & NGC 2005 \\
0.08 & 18.4-18.7 & 1.5 & all \\
0.10 & 18.3 & 1.5 & NGC 2005 \\
0.10 & 18.4-18.7 & 1.5 & all \\
0.12 & 18.3 & 1.5 & NGC 2005 \\
0.12 & 18.4-18.7 & 1.5 & all \\
0.02 & 18.4-18.7 & 2.0 & NGC 1835 \\
0.04 & 18.3 & 2.0 & NGC 2005 \\
0.04 & 18.4-18.7 & 2.0 & all \\
0.06 & 18.3 & 2.0 & NGC 2005 \\
0.06 & 18.4-18.7 & 2.0 & all \\
0.08 & 18.3 & 2.0 & NGC 2005 \\
0.08 & 18.4-18.7 & 2.0 & all \\
0.10 & 18.3 & 2.0 & NGC 2005 \\
0.10 & 18.4-18.7 & 2.0 & all \\
0.12 & 18.3 & 2.0 & NGC 2005 \\
0.12 & 18.4-18.7 & 2.0 & all \\
0.02 & 18.4-18.7 & 2.35 & NGC 1835 \\
0.04 & 18.3 & 2.35 & NGC 2005 \\
0.04 & 18.4-18.7 & 2.35 & all \\
0.06 & 18.3 & 2.35 & NGC 2005 \\
0.06 & 18.4-18.7 & 2.35 & all \\
0.08 & 18.3 & 2.35 & NGC 2005 \\
0.08 & 18.4-18.7 & 2.35 & all \\
0.10 & 18.3 & 2.35 & NGC 2005 \\
0.10 & 18.4-18.7 & 2.35 & all \\
0.12 & 18.3 & 2.35 & NGC 2005 \\
0.12 & 18.4-18.7 & 2.35 & all \\
0.02 & 18.4-18.7 & 3.0 & NGC 1835 \\
0.04 & 18.3 & 3.0 & NGC 2005 \\
0.04 & 18.4-18.7 & 3.0 & all \\
0.06 & 18.3 & 3.0 & NGC 2005 \\
0.06 & 18.4-18.7 & 3.0 & all \\
0.08 & 18.3 & 3.0 & NGC 2005 \\
0.08 & 18.4-18.7 & 3.0 & all \\
0.10 & 18.3 & 3.0 & NGC 2005 \\
0.10 & 18.4-18.7 & 3.0 & all \\
0.12 & 18.3 & 3.0 & NGC 2005 \\
0.12 & 18.4-18.7 & 3.0 & all \\
\enddata
\tablecomments{An age-metallicity relation of $[{\rm Fe/H}] = -0.073 (\frac{Age}{10^9 yr}) - 0.3$
was used for all models}
\label{mpars}
\end{deluxetable}

\clearpage
\begin{deluxetable}{cccccc}
\small
\tablewidth{0pt}
\tablecaption{Parameters of Test Star Formation History}
\tablehead{
\colhead{$E(B-V)$} & 
\colhead{$(m-M)_\circ$} &
\colhead{IMF $x$} &
\colhead{\# Stars bin$^{-1}$} &
\colhead{\# Stars bin$^{-1}$ after} &
\colhead{\# Runs} \\
& & & & completeness correction & \\
}
\startdata
0.08 & 18.5 & 2.35 & 250 & $\sim$185 & 10 \\
0.08 & 18.5 & 2.35 & 500 & $\sim$370 & 10 \\
0.08 & 18.5 & 2.35 & 1000 & $\sim$740 & 25 \\
0.08 & 18.5 & 2.35 & 2000 & $\sim$1480 & 10 \\
0.08 & 18.5 & 2.35 & 3000 & $\sim$2220 & 10 \\
\enddata

\label{tpars}
\end{deluxetable}

\clearpage
\begin{deluxetable}{lcc}
\tablewidth{0pt}
\tablecaption{Best-fit Reddenings and Distances}
\tablehead{
\colhead{Field} &
\colhead{$E(B-V)$} &
\colhead{$(m-M)_\circ$} \\
}
\startdata
NGC 1754 & 0.06 $\pm$ 0.01 & 18.50 $\pm$ 0.05\\
NGC 1835 & 0.04 $\pm$ 0.01 & 18.50 $\pm$ 0.05 \\ 
NGC 1898 & 0.06 $\pm$ 0.01 & 18.40 $\pm$ 0.05 \\ 
NGC 2005 & 0.06 $\pm$ 0.01 & 18.45 $\pm$ 0.05 \\ 
NGC 2019 & 0.06 $\pm$ 0.01 & 18.50 $\pm$ 0.05 \\
\enddata

\label{ebv_dm}
\end{deluxetable}

\clearpage

\clearpage
\epsscale{0.8}

\clearpage
\epsscale{0.7}

\clearpage
\epsscale{0.7}

\clearpage
\epsscale{1.0}

\clearpage
\epsscale{1.0}

\clearpage

\clearpage

\clearpage

\clearpage

\clearpage
\epsscale{0.6}

\clearpage
\epsscale{1.0}

\clearpage
\begin{figure}
\plotone{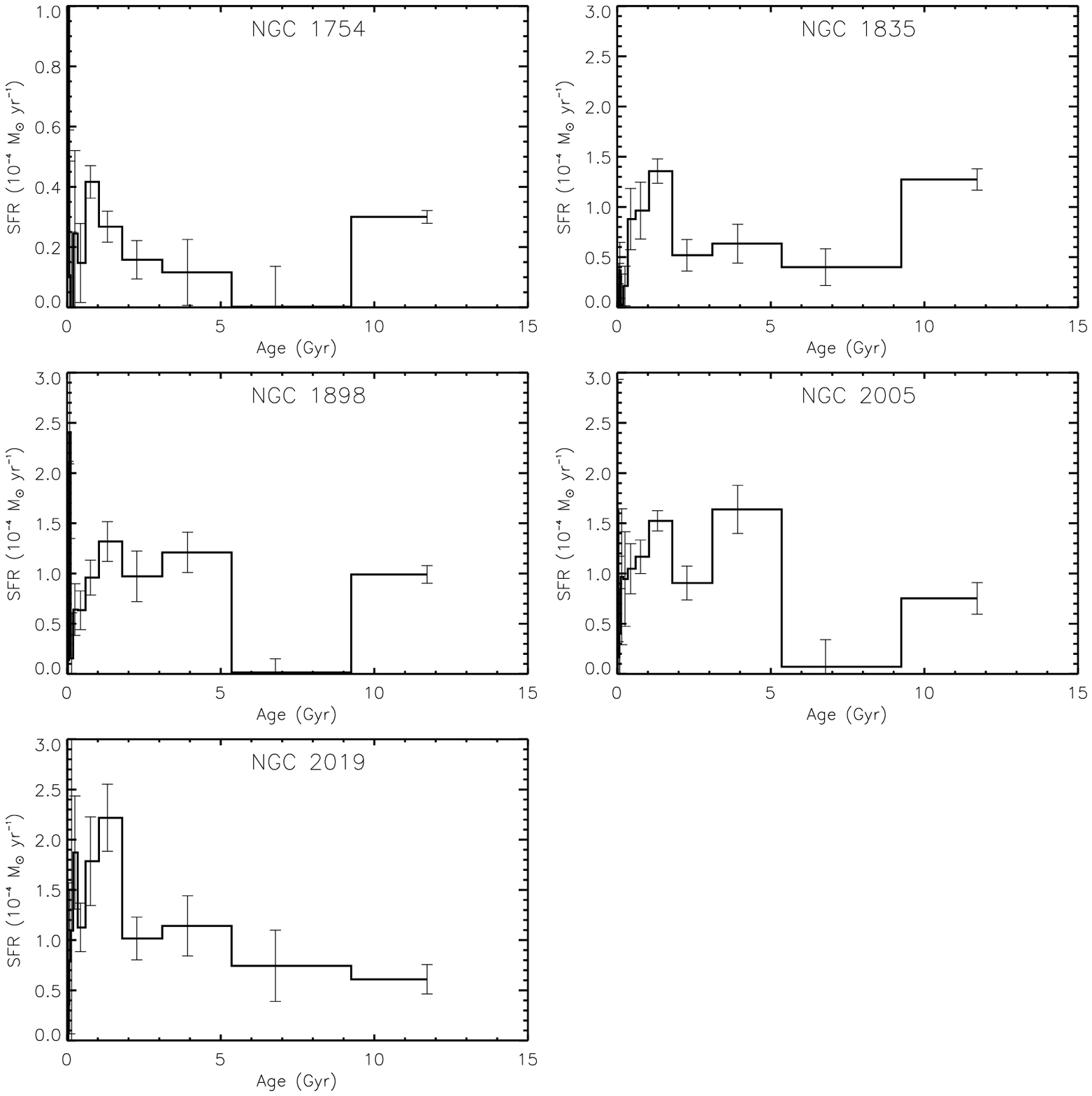}
\end{figure}
\begin{figure}
\plotone{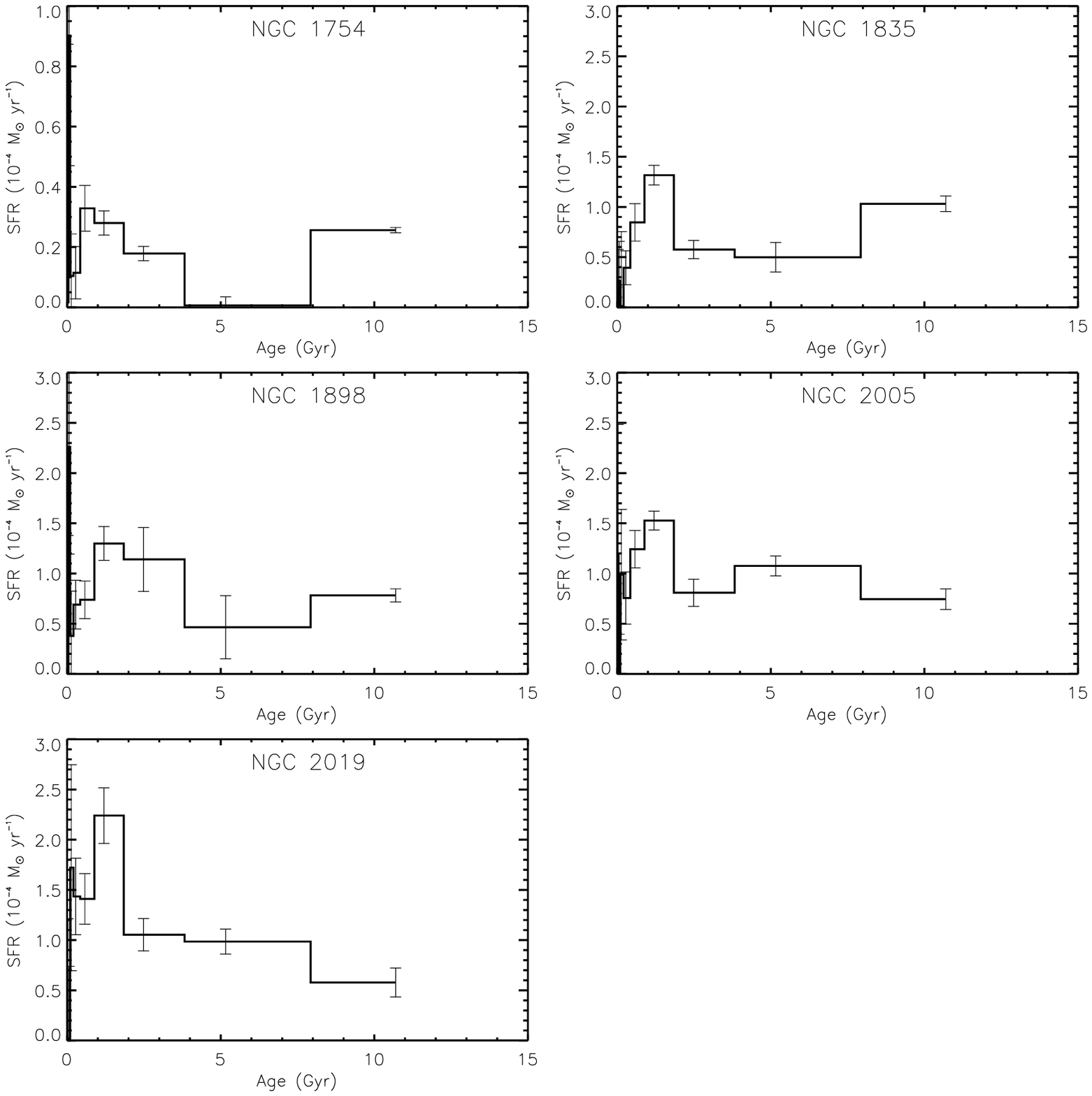}
\end{figure}

\clearpage
\epsscale{0.75}
\begin{figure}
\plotone{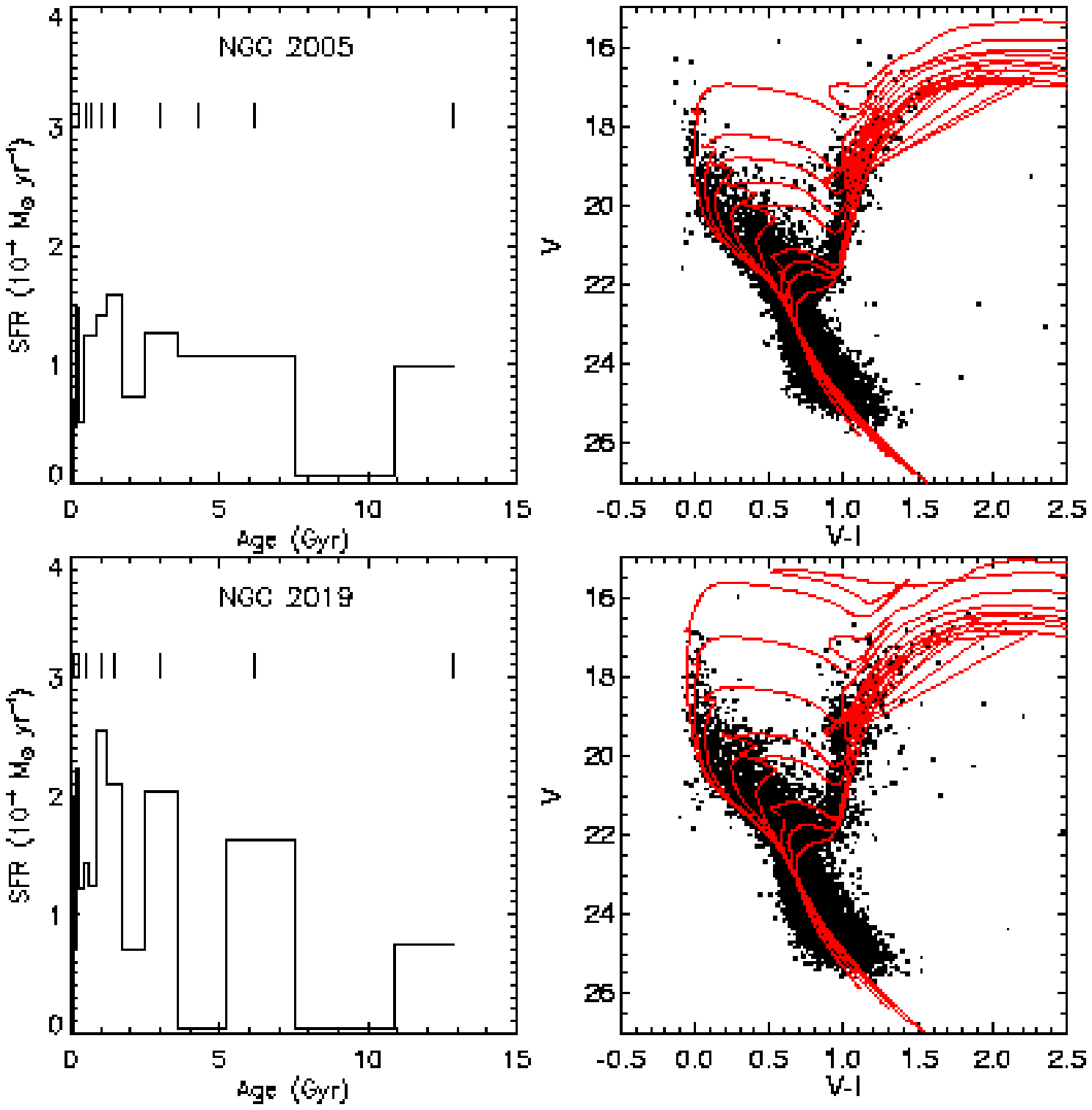}
\end{figure}

\clearpage
\epsscale{0.75}

\clearpage
\epsscale{1.0}

\clearpage
\begin{figure}
\plotone{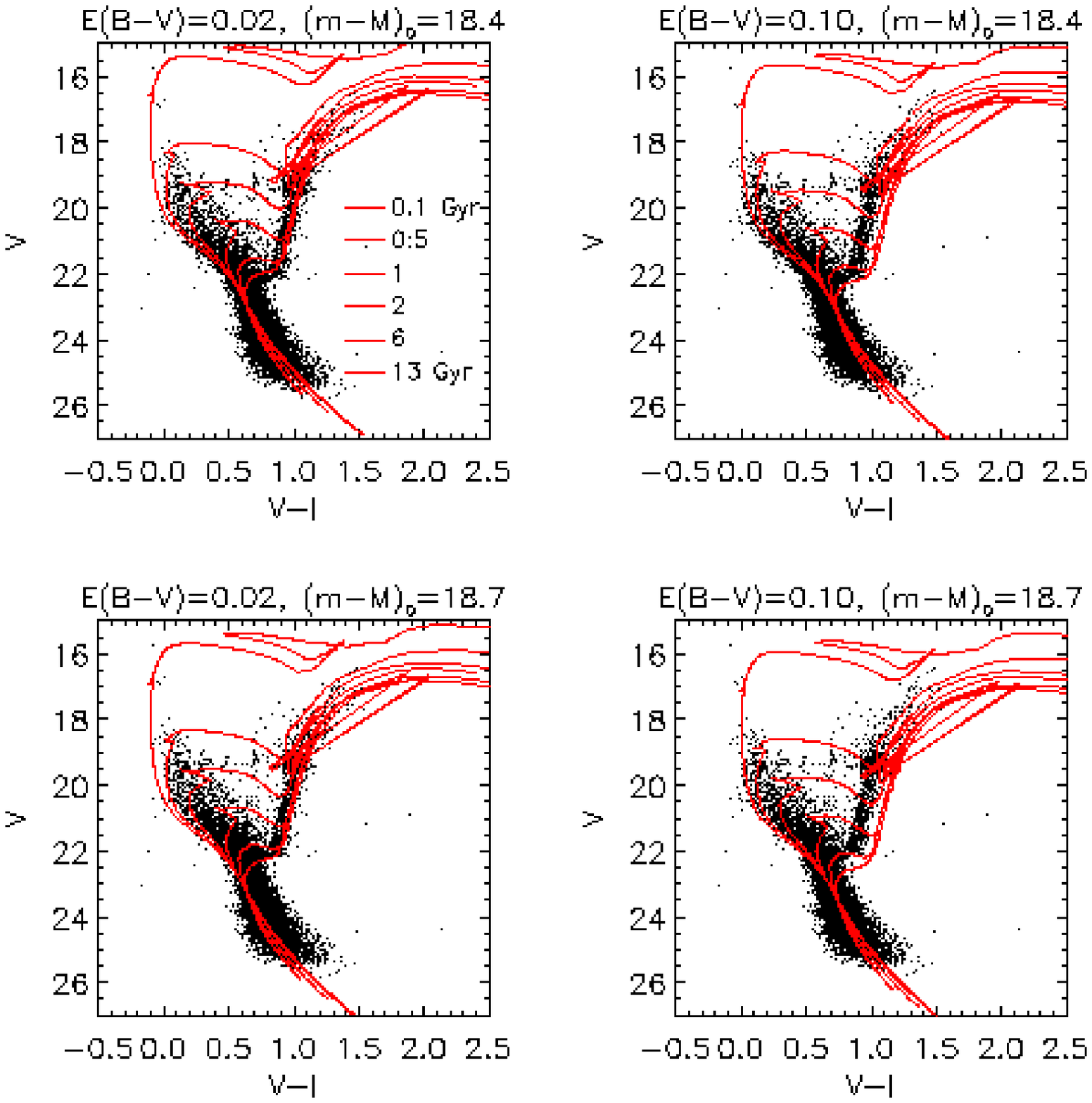}
\end{figure}

\clearpage

\clearpage


\begin{references}
\reference{}Aparicio, A., Gallart, C., \& Bertelli, G. 1997, \aj, 114, 669
\reference{}Ardeberg, A., Gustafsson, B., Linde, P., Nissen, P.-E. 1997, \aap, 322, L13
\reference{}Bertelli, G., Mateo, M., Chiosi, C., \& Bressan, A. 1992, \apj, 387, 320
\reference{}Bertelli, G., Bressan, A., Chiosi, C., Fagotto, F., \& Nasi, E. 1994, A\&A Supp., 106, 275
\reference{}Caldwell, J.A.R., \& Coulson, I.M. 1986, \mnras, 218, 223
\reference{}Chaboyer, B. \& Kim, Y.-C. 1995, \apj, 454, 767
\reference{}Dohm-Palmer, R.C., Skillman, E.D., Gallagher, J., Tolstoy, E., Mateo, M., Dufour, R.J., Saha, A., Hoessel, J.G., \& Chiosi, C. 1998, \aj, 116, 1227
\reference{}Dolphin, A. 1997, New Astronomy, 2, 397
\reference{}Elson, R.A.W., Gilmore, G.F., \& Santiago, B.X. 1997, \mnras, 289, 157
\reference{}Gallagher, J.S. et al.\ 1996, \apj, 466, 732
\reference{}Gallart, C., Aparicio, A., Bertelli, G., \& Chiosi, C. 1996, \aj, 112, 1950
\reference{}Geha, M.C. et al.\ 1998, \aj, 115, 1045
\reference{}Hardy, E., Buonanno, R., Corsi, C.E., Janes, K.A., \& Schommer, R.A. 1984, \apj, 278, 592
\reference{}Holtzman, J. et al.\ 1995, \pasp, 107, 156
\reference{}Holtzman, J.A. et al.\ 1997, \aj, 113, 656
\reference{}Mateo, M., Hodge, P., \& Schommer, R.A. 1986, \apj, 311, 113
\reference{}Ng, Y.K. 1998, A\&A Supp., 132, 133
\reference{} Olsen, K.A.G., Hodge, P.W., Mateo, M., Olszewski, E.W., Schommer, R.A., Suntzeff, N.B., \& Walker, A.R. 1998, \mnras, 300, 665
\reference{}Olszewski, E.W., Schommer, R.A., Suntzeff, N.B., \& Harris, H.C. 1991, \aj, 101, 515
\reference{}Olszewski, E.W., Suntzeff, N.B., \& Mateo, M. 1996, \araa, 34, 511
\reference{}Press, W.H., Teukolsky, S.A., Vetterling, W.T., \& Flannery, B.P. 1992, Numerical Recipes in C, (Cambridge: Cambridge Univ. Press), 2nd edition
\reference{}Rogers, F.J., \& Iglesias, C.A. 1992, \apjs, 79, 507
\reference{}Schaller, G., Schaerer, D., Meynet, G., \& Maeder, A. 1992, A\&A Supp., 96, 269
\reference{}Schechter, P.L., Mateo, M., \& Saha, A. 1993, \pasp, 105, 1342
\reference{}Schwarzschild, M. 1958, Structure and Evolution of the Stars, (New York: Dover), p. 2
\reference{}Smecker-Hane, T. 1997, in AIP Conf. Proc. 393, Star Formation Near and Far, eds. S.S. Holt \& L.G. Mundy, (Woodbury: AIP Press), 571
\reference{}Tolstoy, E., \& Saha, A. 1996, \apj, 462, 672
\reference{}Tolstoy, E., Gallagher, J.S., Cole, A.A., Hoessel, J.G., Saha, A., Dohm-Palmer, R.C., Skillman, E.D., Mateo, M., \& Hurley-Keller, D. 1998, \aj, 116, 1244
\reference{}VandenBerg, D.A. 1998, in IAU Symp. 189, Fundamental Stellar Properties: The Interaction Between Observation and Theory, eds. T.R. Bedding, A.J. Booth, \& J. Davis, (Dordrecht: Kluwer), 439
\reference{}Whitmore, B., \& Heyer, I. 1997, WFPC2 Instrument Science Report 97-08
\end{references}
\end{document}